\documentclass[pra]{revtex4}
 \usepackage{amssymb} \usepackage{graphicx}
\usepackage[utf8]{inputenc}
\usepackage[english]{babel} 
\usepackage{amsthm}
 \usepackage{amsmath}

\begin{document}
 \title{Choices of spinor inner products on M-theory backgrounds}

\author{\"{U}mit Ertem $^{1}$}
\email{umitertemm@gmail.com}
\author{\"{O}zg\"{u}n S\"{u}temen $^{2}$}
\email{sutemen@ankara.edu.tr}
\author{\"{O}zg\"{u}r A\c{c}{\i}k $^{2}$}
\email{ozacik@science.ankara.edu.tr}
\author{Aytolun \c{C}atalkaya $^{2}$}
\email{aytolun.catalkaya@hotmail.com}

\address{$^{1}$ Astronomer, Diyanet \.{I}\c{s}leri Ba\c{s}kanl{\i}\u{g}{\i}, \"{U}niversiteler Mah.\\
 Dumlup{\i}nar Bul. No:147/H 06800 \c{C}ankaya, Ankara, Turkey\\
$^{2}$ Department of Physics, Ankara University,\\ Faculty of Sciences, 06100, Tando\u gan-Ankara,
Turkey}

\begin{abstract}

M-theory backgrounds in the form of unwarped compactifications with or without fluxes are considered. We construct the bilinear forms of supergravity Killing spinors for different choices of spinor inner products on these backgrounds. The equations satisfied by the bilinear forms and their decompositions into product manifolds are obtained for different inner product choices in the special case for which the spinors factorize. It is found that the $AdS$ solutions can only appear for some special choices of spinor inner products on product manifolds. The reduction of bilinears of supergravity Killing spinors into the hidden symmetries of product manifolds which are Killing-Yano and closed conformal Killing-Yano forms for $AdS$ solutions is shown. These hidden symmetries are lifted to eleven-dimensional backgrounds to find the hidden symmetires on them. The relation between the choices of spinor inner products, $AdS$ solutions and hidden symmetries on M-theory backgrounds are investigated.\\
\\
Keywords: Supergravity, flux compactifications, spinor inner products, hidden symmetries

\end{abstract}

\maketitle

\section{Introduction}

Low energy limits of ten-dimensional string theories and eleven-dimensional M-theory correspond to the supergravity theories in those dimensions. The solutions of the bosonic sector of supergravity theories determine the backgrounds that strings and branes can propagate. A special class of backgrounds are obtained by considering compactifications into smaller dimensions in the presence or absence of fluxes defined in relevant supergravity theories \cite{Grana, Douglas Kachru, Becker Becker Schwarz, Curio Kors Lust, Bilal Derendinger Sfetsos, Behrndt Jeschek1, Kaste Minasian Tomasiello, Behrndt Jeschek2, Lukas Saffin, House Micu}. One can consider warped or unwarped products for compactifications and determine the field equations on the product manifolds. The supersymmetry parameters in different bosonic supergravity theories satisfy various supergravity Killing spinor equations arising from the variation of the gravitino field. The bilinear forms of these supergravity Killing spinors can be constructed by using inner products on the spinor space and these bilinears are used in the classification of string and M-theory backgrounds \cite{Gauntlett Pakis, Gauntlett Gutowski Pakis, Gauntlett Gutowski Hull Pakis Reall, Gutowski Martelli Reall, Caldarelli Klemm, Martelli Sparks, DallAgata Prezas, Gauntlett Martelli Waldram, Gauntlett Martelli Pakis Waldram, Saffin, Freedman VanProeyen, DAuria Ferrara Lledo Varadarajan}. Moreover, these bilinears can have Lie algebra structures in some special cases \cite{Acik Ertem}. However, one can define various spinor inner products depending on the dimension and the signature of the background and the corresponding bilinears will be different for different choices of spinor inner products \cite{Acik Ertem2, Acik Ertem1}. In the literature, only some special choices of spinor inner products are considered and there is no exhaustive investigation for all types of inner products and the bilinear forms constructed out of them. Geometric algebra techniques for the construction of bilinear forms in flux backgrounds are considered in \cite{Lazaroiu Babalic Coman1, Lazaroiu Babalic, Lazaroiu Babalic Coman2, Babalic Lazaroiu} which investigate the deeper mathematical structures on supergravity backgrounds. On the other hand, supergravity Killing spinors reduce to geometric Killing spinors or parallel spinors on compactified backgrounds depending on the geometric properties of the product manifolds and the existence of these special types of spinors are related to the special holonomy structures of manifolds \cite{Wang, Berger, Joyce, Ertem}. So, the bilinear forms of supergravity Killing forms can reduce to special types of differential forms on product manifolds and the investigation of these reductions can have implications on the classification problem of string and M-theory backgrounds in all dimensions.

The fermionic terms in the full supergravity Lagrangian are written in terms of spinor inner products. So, the different choices of inner products can also modify the properties of fermionic terms in the Lagrangian. Although the quadratic and quartic fermionic coupling terms will still be nonzero since they contain couplings with bosonic fields, the properties of inner products can change the variational properties of these terms. Hence, the different choices of spinor inner products will lead to different supergravity theories which are equivalent in the bosonic part but different in the fermionic part. This can also change the variation of the gravitino and hence the supersymmetry properties. Supergravity Killing spinor equations contain both the bosonic fields and spinor parameters. Since the change of spinor inner product will change the bilinear properties of spinor parameters, it will lead to different restrictions on the constant terms in the bosonic fields ($\lambda$ and $\mu$ parameters in the 4-form field $F$ given in (8)) due to the changed supergravity Killing form equations and hence lead to the existence or non-existence of non-zero cosmological constant solutions. So, if we insist that the new theory arising from the different choice of the inner product will also be a supersymmetric theory, then the supergravity Killing spinor equations will change in that case. However, if the theory need not be a supersymmetric theory, then we do not need to change the supergravity Killing spinor equations. Our paper considers the effects of the different choices of spinor inner products for a class of eleven-dimensional supergravity theories with common bosonic parts and the same Killing spinor equations. So, the solutions for different spinor inner products can correspond to the solutions of standard physical supergravity theories and also other theories with common bosonic part which are not supersymmetric. Choices of spinor inner products corresponding to physical theories and non-supersymmetric theories are discussed in Appendix A. The case of eleven-dimensional supergravity theories with common bosonic parts and changed supergravity Killing spinor equations due to the choice of spinor inner products which are supersymmetric theories should be investigated in another work.

In this paper, we consider eleven-dimensional M-theory backgrounds in the form of unwarped compactifications with or without fluxes. For the eleven-dimensional background $M_{11}$, the unwarped product structures $M_4\times M_7$, $M_7\times M_4$, $M_5\times M_6$, $M_6\times M_5$ and $M_3\times M_8$ are considered. We determine the decompositions of field equations and supergravity Killing spinor equation onto product manifolds and summarize the possible solutions. We construct bilinear forms of supergravity Killing spinors for both types of spinor inner products on $M_{11}$ and find the equations satisfied by those bilinears. It is found that the non-zero bilinear forms are dependent on the choice of the inner product. We also find the decompositions of the bilinear form equations onto product manifolds which are also highly dependent on the choice of the spinor inner products on product manifolds. Decomposition of bilinear forms on product manifolds will generally correspond to only one choice of spinor inner product on the factors, but in some cases there might exist degenerate situations where both spinor inner products work. $AdS$ solutions will appear in those degenerate situations in which the non-zero flux components can be chosen consistently. The existence of Minkowski or $AdS$ solutions does not depend on the choice of inner products on the background manifold, but the appearence of non-zero flux and hence $AdS$ solutions restricts the possible choices of spinor inner products on product manifolds. So, studying the dependence of the bilinear forms on the spinor inner product gives the consistent inner products for the presence of non-zero flux components while for the fluxless case the inner product choice is irrelevant. Moreover, while the supergravity Killing form bilinears of Minkowski solutions reduce to parallel forms on product manifolds, the bilinears of $AdS$ solutions reduce to special Killing-Yano (KY) or special closed conformal Killing-Yano (CCKY) forms depending on the choices of the spinor inner products. KY forms are antisymmetric generalizations of Killing vector fields to higher degree differential forms and CCKY forms are a subset of antisymmetric generalizations of conformal Killing vector fields to higher degree forms. These special forms are called the hidden symmetries of manifolds. We also obtain KY and CCKY forms of eleven-dimensional backgrounds by lifting the hidden symmetries on product manifolds. So, we determine the relations  between hidden symmetries, $AdS$ solutions and choices of spinor inner products by exhausting all possibilities for spinor inner product choices. This may be considered as a first step of a classification of backgrounds in terms of spinor inner products.

The paper is organized as follows. In Section II, we summarize the equations for the bosonic sector of eleven-dimensional supergravity. Section III deals with $M_4\times M_7$ type backgrounds. We find the decompositions of field equations and supergravity Killing spinor equation and construct the bilinear form equations for both types of spinor inner products with their decompositions onto product manifolds. In Section IV, the same steps are achieved for $M_7\times M_4$ type backgrounds. Section V includes the situation for other types of backgrounds. In Section VI, the relation between hidden symmetries and $AdS$ solutions are summarized and the lifts of hidden symmetries to eleven-dimensional backgrounds are considered. Section VII concludes the paper. There are also three appendices containing the topics of inner product classes of spinor spaces, Clifford algebra conventions and Clifford bracket and KY forms.

\section{Eleven-dimensional supergravity}

Let us consider an eleven-dimensional Lorentzian spin manifold $M_{11}$, with a metric $g$ and a closed 4-form $F$. $F$ is called the flux 4-form and the bosonic sector of the eleven-dimensional supergravity theory defined on $M_{11}$ is given by the following action
\begin{equation}
S=\frac{1}{12\kappa_{11}^2}\int\left(R_{AB}\wedge *_{11}e^{AB}-\frac{1}{2}F\wedge *_{11}F-\frac{1}{6}\mathcal{A}\wedge F\wedge F\right)
\end{equation}
where $\kappa_{11}$ is the eleven-dimensional gravitational coupling constant, capital letter indices take values $A, B=0,1,2,...,9,10$ and $*_{11}$ is the eleven-dimensional Hodge star operator. $R_{AB}$ are the curvature 2-forms, $e^{A}$ are co-frame basis and $\mathcal{A}$ is the 3-form potential of the flux 4-form $F=d\mathcal{A}$. The first term in (1) corresponds to the gravitational term and second and third terms are Maxwell-like and Chern-Simons terms, respectively. The field equations of the eleven-dimensional bosonic supergravity results from the above action by considering the variations of $e^{A}$ and $\mathcal{A}$ as follows
\begin{eqnarray}
*_{11}(i_{X_B}P_A)&=&\frac{1}{2}i_{X_A}F\wedge *_{11}i_{X_B}F-\frac{1}{6}g_{AB}F\wedge *_{11}F\quad\quad\text{(Einstein)}\\
d*_{11}F&=&\frac{1}{2}F\wedge F\quad\quad\quad\quad\quad\quad\quad\quad\quad\quad\quad\quad\quad\quad\,\,\text{(Maxwell)}\\
dF&=&0\quad\quad\quad\quad\quad\quad\quad\quad\quad\quad\quad\quad\quad\quad\quad\quad\quad\,\text{(Closure)}
\end{eqnarray}
where $i_{X_A}$ denotes the interior derivative or contraction operator with respect to the vector field $X_A$, $g_{AB}$ are components of the metric and $P_{A}$ are the Ricci 1-forms defined from the curvature 2-forms as $P_{A}=i_{X^B}R_{BA}$. The last equation (4) is the integrability condition for the definition of the flux form $F$. Moreover, the variation of the gravitino field in the fermionic sector will also lead to a condition on the spinor $\epsilon$ which is the supersymmetry parameter and in the bosonic sector it gives the following supergravity Killing spinor equation
\begin{equation}
\nabla_{X^A}\epsilon=-\frac{1}{24}\left(e^A.F-3F.e^A\right).\epsilon
\end{equation}
where $\nabla_{X^A}$ corresponds to the spinor covariant derivative and $.$ denotes the Clifford multiplication. The co-frame basis $e^A$ define a basis of the Clifford algebra bundle $Cl_{10,1}$ on $M_{11}$ with the following equality
\begin{eqnarray}
e^A.e^B+e^B.e^A=2g^{AB}
\end{eqnarray}
where $g^{AB}$ are the components of the inverse metric. The supersymmetry parameter $\epsilon$ is an element of the spinor bundle $S$ which corresponds to $\mathbb{R}^{32}$ on $M_{11}$ and hence $\epsilon$ is a Majorana spinor.

In the following chapters, we consider various types of unwarped compactifications of supergravity backgrounds which are the solutions of the field equations (2)-(4) of type $M=M_d\times M_{11-d}$. By constructing the bilinear forms of supergravity Killing spinors defined in (5) in those backgrounds, we show that the reduction or non-reduction of those bilinear forms into KY and CCKY forms on product manifolds require the existence or non-existence of $AdS$ or Minkowski type solutions with or without internal and external fluxes. Moreover, we determine the correspondences between the choices of spinor inner products on product manifolds and the types of possible supergravity backgrounds. This gives a classification of unwarped compactifications of supergravity backgrounds in terms of spinor inner products.

\section{$M_4\times M_7$ type backgrounds}

We first consider the case that the eleven-dimensional supergravity background $M_{11}$ has the product structure $M_{11}=M_4\times M_7$ where $M_4$ is a Lorentzian spin 4-manifold and $M_7$ is a Riemannian spin 7-manifold. The frame and co-frame basis indices appeared in the previous equations will split into two parts $A=\{a,\alpha\}$ with $a=0,1,2,3$ and $\alpha=4,5,...,9,10$. The Clifford algebra basis $e^A$ will decompose as
\begin{equation}
e^A=\{e^a\otimes 1_7, iz_4\otimes e^{\alpha}\}
\end{equation}
where $e^a$ are the Clifford algebra basis on $M_4$, $z_4$ is the volume form on $M_4$, $1_7$ is the identity on $M_7$ and $e^{\alpha}$ are the Clifford algebra basis on $M_7$ which are pure imaginary. By considering the equalities $e^a.e^b+e^b.e^a=2g^{ab}$, $e^{\alpha}.e^{\beta}+e^{\beta}.e^{\alpha}=2g^{\alpha\beta}$ and the properties $z_4^2=-1$ and $z_4$ anticommutes with all 1-forms on $M_4$, one can obtain the defining relation (6) from (7). Here, the inverse metric is decomposed as $g^{AB}=\{g^{ab}, g^{\alpha\beta}\}$ and there is no warped product factor. Similarly, the flux 4-form $F$ will decompose as
\begin{equation}
F=\{\lambda iz_4, \mu\phi\}
\end{equation}
where $\lambda$ and $\mu$ are constants and $\phi$ is a 4-form on $M_7$. The flux components on $M_4$ and $M_7$ are called external and internal fluxes respectively and the constants $\lambda$ and $\mu$ determine the existence or non-existence of external and internal flux components. The supersymmetry parameter $\epsilon$ will be constructed from four-dimensional and seven-dimensional spinors $\epsilon_4$ and $\epsilon_7$ as
\begin{equation}
\epsilon=\epsilon_4\otimes\epsilon_7.
\end{equation}

For the product structure $M_4\times M_7$, the field equations (2)-(4) will decompose into four-dimensional and seven-dimensional equations. For the Maxwell equation (3) and the closure condition (4), we can use the decomposition of the flux 4-form $F$ in (8). For any product structure $M_n=M_p\times M_q$, the Hodge star operator $*_n$ satisfies the following equality
\begin{equation}
*_n(\alpha\wedge\beta)=(-1)^{l(p-k)}*_p\alpha\wedge *_q\beta
\end{equation}
where $\alpha$ is a $k$-form on $M_p$ and $\beta$ is a $l$-form on $M_q$ \cite{Alekseevsky Chrysikos Taghavi}. So, in our case, we have
\begin{eqnarray}
*_{11}F&=&i\lambda*_{11}(z_4\wedge 1_7)+\mu *_{11}(1_4\wedge\phi)\nonumber\\
&=&i\lambda(*_4z_4\wedge *_71_7)+\mu(*_41_4\wedge *_7\phi)\nonumber\\
&=&-i\lambda z_7+\mu z_4\wedge *_7\phi
\end{eqnarray}
where we have used $z_4=*_41_4$, $*_4*_4=-1$ and $z_7=*_71_7$. Its exterior derivative gives
\begin{equation}
d*_{11}F=\mu z_4\wedge d*_7\phi
\end{equation}
and the right hand side of (3) is
\begin{equation}
F\wedge F=2i\lambda\mu z_4\wedge\phi
\end{equation}
since we have $dz_4=0$, $z_4\wedge z_4=0$ and $\phi\wedge\phi=0$ because of the fact that $\phi\wedge\phi$ is a 8-form on $M_7$. On the other hand, the closure condition $dF=0$ gives $d\phi=0$ and hence we obtain the following equalities from equations (3) and (4)
\begin{eqnarray}
d*_7\phi&=&i\lambda\phi\nonumber\\
d\phi&=&0.
\end{eqnarray}
These equalities define a weak $G_2$ structure on $M_7$ provided that $\phi$ is a stable 4-form as defined in \cite{Hitchin}. Then, $\phi$ corresponds to the coassociative 4-form of the weak $G_2$ structure \cite{Alekseevsky Chrysikos Taghavi, Friedrich Kath Moroianu Semmelmann}. So, $M_7$ will correspond to a proper weak $G_2$ manifold, a Sasaki-Einstein manifold or a 3-Sasaki manifold \cite{Friedrich Kath Moroianu Semmelmann}. Moreover, (14) means that $\phi$ is a special CCKY 4-form on $M_7$ and hence it must be generated from a geometric Killing spinor \cite{Acik Ertem1}. Since $z_4$ is the volume form on $M_4$, it corresponds to a KY form on $M_4$ and as a result, the flux 4-form $F$ in (8) is generated by KY and CCKY forms on $M_4$ and $M_7$ for $\lambda\neq 0$ and $\mu\neq 0$.

Einstein field equations given in (2) can also be decomposed into $M_4$ and $M_7$ components. From the flux 4-form $F$ in (8), one can find the terms on the right hand side of (2) with similar calculations to the above as follows
\begin{eqnarray}
F\wedge *_{11}F&=&\left(\lambda^2+\mu^2g_4(\phi, \phi)\right)z_{11}\\
i_{X_A}F\wedge *_{11}i_{X_B}F&=&\{-\lambda^2g_{ab}z_{11}, \mu^2g_4(\phi, \phi)g_{\alpha\beta}z_{11}\}
\end{eqnarray}
where $g_p$ denotes the metric on $p$-forms. Here, we have used the definition of Hodge star in terms of the $p$-form metric; for any $p$-forms $\alpha$ and $\beta$ we have $\alpha\wedge *\beta=g_p(\alpha, \beta)*1$. Hence, we have
\begin{eqnarray}
\phi\wedge *_7\phi&=&g_4(\phi, \phi)z_7\nonumber\\
&=&\left((i_{X_{\alpha}}i_{X_{\beta}}i_{X_{\gamma}}i_{X_{\delta}}\phi)i_{X^{\alpha}}i_{X^{\beta}}i_{X^{\gamma}}i_{X^{\delta}}\phi\right)z_7.\nonumber
\end{eqnarray}
The left hand side of (2) corresponds to $i_{X_B}P_A=\{i_{X_b}P_a, i_{X_{\beta}}P_{\alpha}\}$ where $P_a$ and $P_{\alpha}$ are Ricci 1-forms on $M_4$ and $M_7$, respectively. So, the Einstein field equations decompose into $M_4$ and $M_7$ as follows 
\begin{eqnarray}
i_{X_b}P_a&=&-\frac{1}{3}\left(2\lambda^2+\frac{\mu^2}{2}g_4(\phi, \phi)\right)g_{ab}\\
i_{X_{\beta}}P_{\alpha}&=&-\frac{1}{6}\left(\lambda^2+\mu^2g_4(\phi, \phi)\right)g_{\alpha\beta}+\frac{\mu^2}{2}g_3(i_{X_{\alpha}}\phi, i_{X_{\beta}}\phi).
\end{eqnarray}
This means that for $\lambda=\mu=0$, both $M_4$ and $M_7$ are Ricci-flat manifolds and for the special case of $\lambda\neq0$ and $\mu=0$, $M_4$ is a negative curvature and $M_7$ is a positive curvature Einstein manifolds (since the basis 1-forms are pure imaginary on $M_7$, the metric components $g_{\alpha\beta}=g(e_{\alpha}, e_{\beta})$ will have an extra minus sign).

We will also analyze the decomposition of supergravity Killing spinor equation (5) into product manifolds. From (9), the left hand side of (5) corresponds to
\[
\nabla_{X^A}\epsilon=\nabla_{X^a}\epsilon_4\otimes\epsilon_7+\epsilon_4\otimes\nabla_{X^{\alpha}}\epsilon_7
\]
and by using the decompositions in (7) and (8), the right hand side of (5) gives
\begin{eqnarray}
(e^A.F-3F.e^A).\epsilon&=&i\lambda(e^a.z_4-3z_4.e^a).\epsilon_4\otimes\epsilon_7-2e^a.\epsilon_4\otimes\mu\phi.\epsilon_7\nonumber\\
&&-2\lambda\epsilon_4\otimes e^{\alpha}.\epsilon_7+iz_4.\epsilon_4\otimes\mu(e^{\alpha}.\phi-3\phi.e^{\alpha}).\epsilon_7.
\end{eqnarray}
So, the supergravity Killing spinor equation can be written as
\begin{eqnarray}
\nabla_{X^a}\epsilon_4\otimes\epsilon_7+\epsilon_4\otimes\nabla_{X^{\alpha}}\epsilon_7&=&\mp\frac{1}{6}\lambda e^a.\epsilon_4\otimes\epsilon_7+\frac{1}{12}e^a.\epsilon_4\otimes\mu\phi.\epsilon_7\nonumber\\
&&+\frac{1}{12}\lambda\epsilon_4\otimes e^{\alpha}.\epsilon_7\mp\frac{1}{24}\epsilon_4\otimes\mu(e^{\alpha}.\phi-3\phi.e^{\alpha}).\epsilon_7
\end{eqnarray}
where we have used that the volume form $z_4$ anticommutes with basis 1-forms on even dimensions that is $z_4.e^a=-e^a.z_4$ and on a Lorentzian 4-manifold it satisfies $(iz_4)^2=1$, and we also require $iz_4.\epsilon_4=\pm\epsilon_4$. The decompositions of supergravity Killing spinor equation on $M_4$ and $M_7$ have to be considered separately for the cases of existence or nonexistence of internal and external fluxes. For the fluxless case $\lambda=\mu=0$, we have
\begin{eqnarray}
\nabla_{X^a}\epsilon_4=0\nonumber\\
\nabla_{X^{\alpha}}\epsilon_7=0
\end{eqnarray}
and this means that $\epsilon_4$ and $\epsilon_7$ are parallel spinors on $M_4$ and $M_7$, respectively. This is consistent with the Ricci-flatness property in (17) and (18). 7-dimensional Riemannian manifolds admitting parallel spinors correspond to $G_2$ holonomy manifolds \cite{Berger}. 4-dimensional Lorentzian manifolds admitting parallel spinors can be Minkowski or plane-wave spacetimes. However, Ricci-flatness property restricts the case to the Minkowski spacetime. Then, this case corresponds to the solution $\text{Mink}_4\times G_2$. For the existence of only the external flux $\lambda\neq0$ and $\mu=0$, we have
\begin{eqnarray}
\nabla_{X^a}\epsilon_4&=&\mp\frac{1}{6}\lambda e^a.\epsilon_4\nonumber\\
\nabla_{X^{\alpha}}\epsilon_7&=&-\frac{1}{12}\lambda e^{\alpha}.\epsilon_7
\end{eqnarray}
and this corresponds to the case that $\epsilon_4$ and $\epsilon_7$ are geometric Killing spinors on $M_4$ and $M_7$ respectively which is consistent with being Einstein manifolds from (17) and (18). The geometric Killing spinors on $M_4$ and $M_7$ are real and imaginary Killing spinors, respectively \footnote{In some papers, the sign convention for the Clifford algebra is chosen as $e^a.e^b+e^b.e^a=-2g^{ab}$ and the real and imaginary Killing spinors appear on $M_7$ and $M_4$, respectively which is reverse to our sign convention $e^a.e^b+e^b.e^a=2g^{ab}$ which gives real and imaginary Killing spinors on $M_4$ and $M_7$, respectively.}. 7-dimensional Riemannian manifolds admitting imaginary Killing spinors correspond to weak $G_2$ manifolds. In the case of admitting one Killing spinor, it is a proper weak $G_2$ manifold. For the existence of two and three Killing spinors, it corresponds to Sasaki-Einstein and 3-Sasaki manifolds, respectively. If there are maximal number of Killing spinors, then $M_7$ is a round sphere $S^7$. 4-dimensional Einstein manifolds with negative curvature admitting real Killing spinors correspond to $AdS_4$ spacetimes. Then, the solutions in that case corresponds to $AdS_4\times S^7$ and $AdS_4\times\text{weak }G_2$. For the general case of $\lambda\neq0$ and $\mu\neq0$, we have both nonzero internal and external fluxes. In the literature, the presence of internal fluxes generally implies the consideration of a warp factor in the metric to obtain consistent solutions of field equations \cite{Behrndt Jeschek1, Kaste Minasian Tomasiello, Behrndt Jeschek2, Lukas Saffin, House Micu, Fre Trigiante}. However, there is also a possibility of a solution for the unwarped case if the internal flux component $\phi$ satisfies a specific condition. We know that the internal flux $\phi$ satisfies the special CCKY form equations (14) which means that they are constructed from geometric Killing spinors. If $\phi$ satisfies the condition $\phi.\epsilon_7=\pm\frac{1}{2}\epsilon_7$, then the supergravity Killing spinor equation decomposes into the following equations
\begin{eqnarray}
\nabla_{X^a}\epsilon_4&=&-\frac{1}{6}\left(\pm\lambda+\frac{\mu}{4}\right)e^a.\epsilon_4\\
\nabla_{X^{\alpha}}\epsilon_7&=&\frac{1}{12}\left(\lambda\pm\frac{\mu}{4}\right)e^{\alpha}.\epsilon_7\pm\frac{\mu}{8}\phi.e^{\alpha}.\epsilon_7.
\end{eqnarray}
Moreover, one can write the Clifford product of a 1-form $e^{\alpha}$ with an arbitrary form $\omega$ in terms of the wedge product and interior derivative as follows
\begin{eqnarray}
e^{\alpha}.\omega&=&e^{\alpha}\wedge\omega+i_{X^{\alpha}}\omega\nonumber\\
\omega.e^{\alpha}&=&e^{\alpha}\wedge\eta\omega-i_{X^{\alpha}}\eta\omega
\end{eqnarray}
where the automorphism $\eta$ acts on a $p$-form $\omega$ as $\eta\omega=(-1)^p\omega$. Then, we have
\begin{equation}
\phi.e^{\alpha}=e^{\alpha}.\phi-2i_{X^{\alpha}}\phi.
\end{equation}
By applying the interior derivative operator $i_{X^{\alpha}}$ to the equations (14), one can see that $\phi$ satisfies
\begin{eqnarray}
di_{X^{\alpha}}*_7\phi&=&-\frac{3i\lambda}{4}i_{X^{\alpha}}\phi\nonumber\\
i_{X^{\alpha}}d*_7\phi&=&i\lambda i_{X^{\alpha}}\phi
\end{eqnarray}
and from the definition of the Lie derivative $\mathcal{L}_{X^{\alpha}}=di_{X^{\alpha}}+i_{X^{\alpha}}d$ on forms, one obtains
\begin{equation}
\mathcal{L}_{X^{\alpha}}*_7\phi=\frac{i\lambda}{4}i_{X^{\alpha}}\phi.
\end{equation}
If the following condition on $\phi$ is satisfied
\begin{equation}
(\mathcal{L}_{X^{\alpha}}*_7\phi).\epsilon_7=i\lambda e^{\alpha}.\epsilon_7
\end{equation}
then the equation (24) is transformed into
\begin{eqnarray}
\nabla_{X^{\alpha}}\epsilon_7=\frac{1}{12}\left(\lambda\mp\frac{25}{2}\mu\right)e^{\alpha}.\epsilon_7.
\end{eqnarray}
Now, if we choose the constant $\mu$ as $\mu=\pm\frac{\lambda}{5}$, then the supergravity Killing spinor equation decomposes into the following equations from (23) and (30)
\begin{eqnarray}
\nabla_{X^a}\epsilon_4&=&\mp\frac{7}{40}\lambda e^a.\epsilon_4\\
\nabla_{X^{\alpha}}\epsilon_7&=&-\frac{1}{8}\lambda e^{\alpha}.\epsilon_7
\end{eqnarray}
which correspond to geometric Killing spinors on $M_4$ and $M_7$. This is consistent with the condition $\phi.\epsilon_7=\pm\frac{1}{2}\epsilon_7$, since if $\phi$ is constructed from $\epsilon_7$ as a bilinear 4-form, then it automatically satisifies this condition from Fierz identities \cite{Acik Ertem1}. So, the only restriction on $\phi$ to obtain geometric Killing spinors on product manifolds is the condition (28). Indeed, the equations satisfied by $\phi$ correspond to the case that $M_7$ is a weak $G_2$ manifold and $\phi$ is the coassociative 4-form defined on it. In that case, $g_4(\phi, \phi)$ in (17) and (18) is constant and $g_3(i_{X_{\alpha}}\phi, i_{X_{\beta}}\phi)$ is proportional to $g_{\alpha\beta}$ \cite{Bryant}. So, equations (17) and (18) imply that $M_4$ and $M_7$ are Einstein manifolds. Then, the case $\lambda\neq0$ and $\mu\neq0$, for the special choice of $\mu=\pm\frac{\lambda}{5}$, also corresponds to the solutions $AdS_4\times S^7$ and $AdS_4\times \text{weak }G_2$. But, for the general case of $\lambda\neq0$ and $\mu\neq0$, the supergravity Killing spinor equation (20) cannot be decomposed into $M_4$ and $M_7$ components and one cannot find a general solution. For the final case of $\lambda=0$ and $\mu\neq0$ which corresponds to the existence of only the internal flux, the equations will be similar to the previous case. However, if we take $\lambda=0$ in (17), (18), (31) and (32), then (17) and (18) imply that $M_4$ and $M_7$ are Einstein manifolds, but (31) and (32) imply that they must admit parallel spinors which is inconsistent. So, $\lambda=0$ and $\mu\neq0$ case does not correspond to a solution.

\subsection{Bilinear forms}

Now, we will construct bilinear forms of supergravity Killing spinors by using the defining equation (5). The spinor bilinear of a spinor $\epsilon$ is defined in terms of the spinor inner product $(\,,\,)$ and co-frame basis as a sum of different degree differential forms as follows
\[
\epsilon\overline{\epsilon}=(\epsilon, \epsilon)+(\epsilon, e_a.\epsilon)e^a+(\epsilon, e_{ba}.\epsilon)e^{ab}+...+(\epsilon, e_{a_p...a_2a_1}.\epsilon)e^{a_1a_2...a_p}+...+(-1)^{\lfloor n/2\rfloor}(\epsilon, z.\epsilon)z
\]
where $e^{a_1a_2...a_p}=e^{a_1}\wedge e^{a_2}\wedge ...\wedge e^{a_p}$ and $z$ is the volume form. The bilinear $p$-form of the spinor $\epsilon$ is defined as the $p$-form component of the spinor bilinear
\begin{equation}
(\epsilon\overline{\epsilon})_p=(\epsilon, e_{a_p...a_2a_1}.\epsilon)e^{a_1a_2...a_p}.
\end{equation}
However, we can consider two different spinor inner products on the spinor bundle of $M_{11}$. We have the Clifford algebra $Cl_{10,1}$ and its even subalgebra that is isomorphic to $Cl^0_{10,1}\cong Cl_{1,9}$. So, the spinor space is isomorphic to $\mathbb{R}^{32}$ and we have Majorana spinors with the spinor inner product choices $\mathbb{R}$-skew with $\xi\eta$ involution or $\mathbb{R}$-symmetric with $\xi$ involution, where $\xi$ is acting on a $p$-form $\omega$ as $\omega^{\xi}=(-1)^{\lfloor p/2\rfloor}\omega$ with $\lfloor\,\rfloor$ is the floor function. The details of the spinor inner product classes in all dimensions can be found in Appendix A. In the literature, only the first choice of the inner product is considered and the investigations are based on this choice. We will consider both choices separately and analyze the decomposition of bilinear forms on product manifolds in both cases.

\subsubsection{$\mathbb{R}$-skew $\xi\eta$ inner product}

First, we choose the spinor inner product as $\mathbb{R}$-skew with $\xi\eta$ involution and find the decomposition of bilinear forms on product manifolds. The bilinear forms constructed from a spinor are elements of $S\otimes S^*$ where $S$ is the spinor space and $S^*$ is the dual spinor space. Since the connection $\nabla$ is compatible with  the spinor inner product $(\,,\,)$ and preserves the degree of a form, it is also compatible with the projection operation $(\,)_p$ on $p$-form bilinears and we can write for a supergravity Killing spinor $\epsilon$ as
\begin{eqnarray}
\nabla_X(\epsilon\overline{\epsilon})_p&=&\left((\nabla_X\epsilon)\overline{\epsilon}\right)_p+\left(\epsilon\overline{\nabla_X\epsilon}\right)_p\nonumber\\
&=&-\frac{1}{24}\left((\widetilde{X}.F-3F.\widetilde{X}).\epsilon\overline{\epsilon}\right)_p-\frac{1}{24}\left(\epsilon\overline{(\widetilde{X}.F-3F.\widetilde{X}).\epsilon}\right)_p
\end{eqnarray}
where we have used (5). For any spinor $\psi$, the dual spinor $\overline{\psi}$ can be written in terms of the involution operation $\mathcal{J}$ as $\overline{\psi}=\psi^{\mathcal{J}}$. Since we have $\mathcal{J}=\xi\eta$, for any Clifford form $\omega$ and spinor $\psi$, we have $\overline{\omega.\psi}=(\omega.\psi)^{\xi\eta}=\psi^{\xi\eta}.\omega^{\xi\eta}=\overline{\psi}.\omega^{\xi\eta}$. Then, we can write
\begin{eqnarray}
\overline{(\widetilde{X}.F-3F.\widetilde{X}).\epsilon}&=&\overline{\epsilon}.(\widetilde{X}.F-3F.\widetilde{X})^{\xi\eta}\nonumber\\
&=&\overline{\epsilon}.(F^{\xi\eta}.\widetilde{X}^{\xi\eta}-3\widetilde{X}^{\xi\eta}.F^{\xi\eta})\nonumber\\
&=&-\overline{\epsilon}.(F.\widetilde{X}-3\widetilde{X}.F)
\end{eqnarray}
where we have used $F^{\xi\eta}=F$ and ${\widetilde{X}}^{\xi\eta}=-\widetilde{X}$. By using this equality in (34), we obtain
\begin{equation}
\nabla_X(\epsilon\overline{\epsilon})_p=-\frac{1}{24}\left((\widetilde{X}.F-3F.\widetilde{X}).\epsilon\overline{\epsilon}\right)_p+\frac{1}{24}\left(\epsilon\overline{\epsilon}.(F.\widetilde{X}-3\widetilde{X}.F)\right)_p.
\end{equation}
If we add and subtract the term $\frac{1}{6}\left(\epsilon\overline{\epsilon}.(F.\widetilde{X}-\widetilde{X}.F)\right)_p$ to the right hand side, we find
\begin{equation}
\nabla_X(\epsilon\overline{\epsilon})_p=-\frac{1}{24}\left((\widetilde{X}.F-3F.\widetilde{X}).\epsilon\overline{\epsilon}\right)_p+\frac{1}{24}\left(\epsilon\overline{\epsilon}.(\widetilde{X}.F-3F.\widetilde{X})\right)_p+\frac{1}{6}\left(\epsilon\overline{\epsilon}.(F.\widetilde{X}-\widetilde{X}.F)\right)_p.
\end{equation}
So, the bilinear form equation of supergravity Killing spinor $\epsilon$ which is also called the supergravity Killing form equation can be written as
\begin{equation}
\nabla_X(\epsilon\overline{\epsilon})_p=-\frac{1}{24}\left([(\widetilde{X}.F-3F.\widetilde{X}), \epsilon\overline{\epsilon}]_{Cl}\right)_p+\frac{1}{6}\left(\epsilon\overline{\epsilon}.[F, \widetilde{X}]_{Cl}\right)_p
\end{equation}
where $[\,,\,]_{Cl}$ denotes the Clifford bracket. Since we can write
\begin{eqnarray}
\widetilde{X}.F&=&\widetilde{X}\wedge F+i_XF\nonumber\\
F.\widetilde{X}&=&\widetilde{X}\wedge F-i_XF
\end{eqnarray}
and so
\begin{eqnarray}
\widetilde{X}.F-3F.\widetilde{X}&=&-2\widetilde{X}\wedge F+4i_XF\nonumber\\
F.\widetilde{X}-\widetilde{X}.F&=&-2i_XF
\end{eqnarray}
the supergravity Killing form equation (38) turns into
\begin{equation}
\nabla_X(\epsilon\overline{\epsilon})_p=\frac{1}{12}\left([\widetilde{X}\wedge F, \epsilon\overline{\epsilon}]_{Cl}\right)_p-\frac{1}{6}\left([i_XF, \epsilon\overline{\epsilon}]_{Cl}\right)_p-\frac{1}{3}\left(\epsilon\overline{\epsilon}.i_XF\right)_p.
\end{equation}
The only non-zero bilinear forms of a spinor on an eleven-dimensional Lorentzian manifold are 1-, 2-, 5-, 6-, 9- and 10-forms as can be seen from Table XVII in Appendix A. So, the spinor bilinear of the supergravity Killing spinor $\epsilon$ is
\begin{equation}
\epsilon\overline{\epsilon}=(\epsilon\overline{\epsilon})_1+(\epsilon\overline{\epsilon})_2+(\epsilon\overline{\epsilon})_5+(\epsilon\overline{\epsilon})_6+(\epsilon\overline{\epsilon})_9+(\epsilon\overline{\epsilon})_{10}.
\end{equation}
We can find the equations satisfied by all of the bilinear forms by considering the definition of the Clifford bracket and projection operation given in (B9). For $p=1$, we have the following equation for the bilinear 1-form $(\epsilon\overline{\epsilon})_1$ from (41)
\begin{equation}
\nabla_{X_A}(\epsilon\overline{\epsilon})_1=\frac{1}{144}F\underset{4}{\wedge}i_{X_A}(\epsilon\overline{\epsilon})_6-\frac{1}{6}i_{X_A}F\underset{2}{\wedge}(\epsilon\overline{\epsilon})_2
\end{equation}
where we have used the definition of the contracted wedge product given in (B8). If we use the definitions of the exterior derivative and coderivative in terms of the covariant derivative as $d=e^A\wedge\nabla_{X_A}$ and $\delta=-i_{X^A}\nabla_{X_A}$ for zero torsion, we obtain
\begin{eqnarray}
d(\epsilon\overline{\epsilon})_1&=&\frac{1}{72}F\underset{4}{\wedge}(\epsilon\overline{\epsilon})_6-\frac{1}{3}F\underset{2}{\wedge}(\epsilon\overline{\epsilon})_2\nonumber\\
\delta(\epsilon\overline{\epsilon})_1&=&0.
\end{eqnarray}
By comparing the equations (43) and (44), one can easily see that $(\epsilon\overline{\epsilon})_1$ satisfies the equation
\begin{equation}
\nabla_{X_A}(\epsilon\overline{\epsilon})_1=\frac{1}{2}i_{X_A}d(\epsilon\overline{\epsilon})_1
\end{equation}
and hence $(\epsilon\overline{\epsilon})_1$ is a KY 1-form. Consequently, the vector field which is metric dual to the 1-form $(\epsilon\overline{\epsilon})_1$ is a Killing vector field. The definition and properties of KY forms can be found in Appendix C. For $p=2$, the bilinear form equation (41) gives
\begin{equation}
\nabla_{X_A}(\epsilon\overline{\epsilon})_2=\frac{1}{36}F\underset{3}{\wedge}i_{X_A}(\epsilon\overline{\epsilon})_5+\frac{1}{144}e_A\wedge(F\underset{4}{\wedge}(\epsilon\overline{\epsilon})_5)-\frac{1}{3}(\epsilon\overline{\epsilon})_1\underset{1}{\wedge}i_{X_A}F+\frac{1}{18}(\epsilon\overline{\epsilon})_5\underset{3}{\wedge}i_{X_A}F
\end{equation}
and the exterior and co-derivatives are
\begin{eqnarray}
d(\epsilon\overline{\epsilon})_2&=&(\epsilon\overline{\epsilon})_1\underset{1}{\wedge}F\nonumber\\
\delta(\epsilon\overline{\epsilon})_2&=&\frac{11}{72}F\underset{4}{\wedge}(\epsilon\overline{\epsilon})_5.
\end{eqnarray}
So, $(\epsilon\overline{\epsilon})_2$ does not satisfy the KY equation. For $p=5$, the bilinear form equation gives
\begin{eqnarray}
\nabla_{X_A}(\epsilon\overline{\epsilon})_5&=&\frac{1}{6}(e_A\wedge F)\underset{1}{\wedge}(\epsilon\overline{\epsilon})_2-\frac{1}{3}i_{X_A}F\wedge(\epsilon\overline{\epsilon})_2+\frac{1}{6}i_{X_A}F\underset{2}{\wedge}(\epsilon\overline{\epsilon})_6\nonumber\\
&&-\frac{1}{36}(e_A\wedge F)\underset{3}{\wedge}(\epsilon\overline{\epsilon})_6+\frac{1}{6}(e_A\wedge F)\underset{5}{\wedge}(\epsilon\overline{\epsilon})_{10}
\end{eqnarray}
where the terms on the right hand side can also be written in a more explicit way by using the identity
\begin{equation}
(\widetilde{X}\wedge F)\underset{k}{\wedge}\alpha=kF\underset{k-1}{\wedge}i_X\alpha+(-1)^k\widetilde{X}\wedge(F\underset{k}{\wedge}\alpha).
\end{equation}
(48) implies
\begin{eqnarray}
d(\epsilon\overline{\epsilon})_5&=&-F\wedge(\epsilon\overline{\epsilon})_2+\frac{1}{24}F\underset{4}{\wedge}(\epsilon\overline{\epsilon})_{10}\nonumber\\
\delta(\epsilon\overline{\epsilon})_5&=&\frac{2}{3}F\underset{1}{\wedge}(\epsilon\overline{\epsilon})_2-\frac{1}{18}(\epsilon\overline{\epsilon})_6\underset{3}{\wedge}F.
\end{eqnarray}
Similarly, for $p=6$, we have
\begin{eqnarray}
\nabla_{X_A}(\epsilon\overline{\epsilon})_6&=&\frac{1}{144}(e_A\wedge F)\underset{4}{\wedge}(\epsilon\overline{\epsilon})_9-\frac{1}{3}(\epsilon\overline{\epsilon})_5\underset{1}{\wedge}i_{X_A}F+\frac{1}{18}(\epsilon\overline{\epsilon})_9\underset{3}{\wedge}i_{X_A}F\nonumber\\
&&+\frac{1}{6}(e_A\wedge F)\wedge(\epsilon\overline{\epsilon})_1-\frac{1}{12}(e_A\wedge F)\underset{2}{\wedge}(\epsilon\overline{\epsilon})_5
\end{eqnarray}
and
\begin{eqnarray}
d(\epsilon\overline{\epsilon})_6&=&\frac{1}{3}F\underset{1}{\wedge}(\epsilon\overline{\epsilon})_5+\frac{1}{6}F\wedge(\epsilon\overline{\epsilon})_1\nonumber\\
\delta(\epsilon\overline{\epsilon})_6&=&-\frac{1}{144}F\underset{4}{\wedge}(\epsilon\overline{\epsilon})_9+\frac{7}{6}F\underset{1}{\wedge}(\epsilon\overline{\epsilon})_1+F\underset{2}{\wedge}(\epsilon\overline{\epsilon})_5.
\end{eqnarray}
5- and 6-form bilinears also do not satisfy the KY form equation. For the case of $p=9$, (41) gives
\begin{equation}
\nabla_{X_A}(\epsilon\overline{\epsilon})_9=\frac{1}{6}(e_A\wedge F)\underset{1}{\wedge}(\epsilon\overline{\epsilon})_6-\frac{1}{36}(e_A\wedge F)\underset{3}{\wedge}(\epsilon\overline{\epsilon})_{10}-\frac{1}{3}i_{X_A}F\wedge(\epsilon\overline{\epsilon})_6+\frac{1}{6}F\underset{3}{\wedge}(\epsilon\overline{\epsilon})_{10}
\end{equation}
and
\begin{eqnarray}
d(\epsilon\overline{\epsilon})_9&=&-\frac{1}{3}\left(F\wedge(\epsilon\overline{\epsilon})_6+F\underset{2}{\wedge}(\epsilon\overline{\epsilon})_{10}\right)\nonumber\\
\delta(\epsilon\overline{\epsilon})_9&=&\frac{1}{6}F\underset{3}{\wedge}(\epsilon\overline{\epsilon})_{10}.
\end{eqnarray}
For $p=10$, we have
\begin{equation}
\nabla_{X_A}(\epsilon\overline{\epsilon})_{10}=-\frac{1}{12}(e_A\wedge F)\underset{2}{\wedge}(\epsilon\overline{\epsilon})_9-\frac{1}{3}(\epsilon\overline{\epsilon})_9\underset{1}{\wedge}i_{X_A}F+\frac{1}{6}(e_A\wedge F)\wedge(\epsilon\overline{\epsilon})_5
\end{equation}
and
\begin{eqnarray}
d(\epsilon\overline{\epsilon})_{10}&=&-\frac{1}{3}F\underset{1}{\wedge}(\epsilon\overline{\epsilon})_9\nonumber\\
\delta(\epsilon\overline{\epsilon})_{10}&=&-\frac{1}{3}\left(F\wedge(\epsilon\overline{\epsilon})_5+F\underset{2}{\wedge}(\epsilon\overline{\epsilon})_9\right).
\end{eqnarray}
So, except the 1-form bilinear, all the higher degree bilinear forms of supergravity Killing spinors do not correspond to KY forms and satisfy different types of equations.

Now, we can consider the decomposition of bilinear forms onto product manifolds $M_4$ and $M_7$. Since the supergravity Killing spinor $\epsilon$ decomposes as in (9), the spinor bilinears decompose as $\epsilon\overline{\epsilon}=\{\epsilon_4\overline{\epsilon_4}, \epsilon_7\overline{\epsilon_7}\}$. By considering the definitions $\epsilon\overline{\epsilon}^{(4)}:=\epsilon_4\overline{\epsilon_4}$ and $\epsilon\overline{\epsilon}^{(7)}:=\epsilon_7\overline{\epsilon_7}$, the $p$-form bilinears on product manifolds correspond to
\begin{equation}
(\epsilon\overline{\epsilon})_p=\{(\epsilon\overline{\epsilon}^{(4)})_p, (\epsilon\overline{\epsilon}^{(7)})_p\}.
\end{equation}
Since, the degree of differential forms cannot be greater than the volume form, from (42) we have
\begin{eqnarray}
\epsilon\overline{\epsilon}^{(4)}&=&(\epsilon\overline{\epsilon}^{(4)})_1+(\epsilon\overline{\epsilon}^{(4)})_2\nonumber\\
\epsilon\overline{\epsilon}^{(7)}&=&(\epsilon\overline{\epsilon}^{(7)})_1+(\epsilon\overline{\epsilon}^{(7)})_2+(\epsilon\overline{\epsilon}^{(7)})_5+(\epsilon\overline{\epsilon}^{(7)})_6.
\end{eqnarray}
Moreover, depending on the spinor inner product choices on $M_4$ and $M_7$, one can determine the properties of nonzero bilinears constructed out of $\epsilon_4$ and $\epsilon_7$. $M_4$ is a Lorentzian manifold, the spinor space corresponds to $\mathbb{C}^2\oplus\mathbb{C}^2$ and the spinors are Dirac-Weyl spinors. $M_7$ is a Riemannian manifold, the spinor space corresponds to $\mathbb{R}^8$ and the spinors are Majorana spinors. So, from Table XVII in Appendix A, we have the bilinears for the chosen inner products given in Table I.

\begin{table}[h]
\centering
\begin{tabular}{c c|c c c c}

$$ & $\text{inner product}$ & $1$ & $2$ & $5$ & $6$\\ \hline
$i)$ & $M_4 : \mathbb{C}^*\text{-sym }\xi$ & $R$ & $I$ & $$ & $$\\
$$ & $M_7 : \mathbb{R}\text{-skew }\xi$ & $\times$ & $\checkmark$ & $\times$ & $\checkmark$\\ \hline
$ii)$ & $M_4 : \mathbb{C}^*\text{-sym }\xi$ & $R$ & $I$ & $$ & $$\\
$$ & $M_7 : \mathbb{R}\text{-sym }\xi\eta$ & $\times$ & $\times$ & $\times$ & $\times$\\ \hline
$iii)$ & $M_4 : \mathbb{C}\text{-skew }\xi\eta$ & $\checkmark$ & $\checkmark$ & $$ & $$\\
$$ & $M_7 : \mathbb{R}\text{-skew }\xi$ & $\times$ & $\checkmark$ & $\times$ & $\checkmark$\\ \hline
$iv)$ & $M_4 : \mathbb{C}\text{-skew }\xi\eta$ & $\checkmark$ & $\checkmark$ & $$ & $$\\
$$ & $M_7 : \mathbb{R}\text{-sym }\xi\eta$ & $\times$ & $\times$ & $\times$ & $\times$\\

\end{tabular}
\caption{Properties of nonzero bilinears for different spinor inner product choices on Lorentzian $M_4$ and Riemannian $M_7$ for $\mathbb{R}$-skew $\xi\eta$ inner product on $M_{11}$.}
\end{table}

So, we have to consider four different cases separately in the decomposition of bilinear forms onto product manifolds.

i) $M_4 : \mathbb{C}^*\text{-sym }\xi$ and $M_7 : \mathbb{R}\text{-skew }\xi$;

By considering the nonzero bilinears in Table I and the decomposition of the 4-form flux $F$ in (8), the 1-form bilinear equation (43) decomposes as
\begin{eqnarray}
\nabla_{X_a}(\epsilon\overline{\epsilon}^{(4)})_1&=&-\frac{i\lambda}{6}i_{X_a}z_4\underset{2}{\wedge}(\epsilon\overline{\epsilon}^{(4)})_2\nonumber\\
0&=&\frac{\mu}{144}\phi\underset{4}{\wedge}i_{X_{\alpha}}(\epsilon\overline{\epsilon}^{(7)})_6-\frac{\mu}{6}i_{X_{\alpha}}\phi\underset{2}{\wedge}(\epsilon\overline{\epsilon}^{(7)})_2
\end{eqnarray}
and the second equality implies that $\mu=0$ while the first equality implies that $\lambda$ is real (since $(\epsilon\overline{\epsilon}^{(4)})_1$ is real and $(\epsilon\overline{\epsilon}^{(4)})_2$ is pure imaginary from Table I). From the first equality, one can obtain
\begin{eqnarray}
d(\epsilon\overline{\epsilon}^{(4)})_1&=&-\frac{i\lambda}{3}z_4\underset{2}{\wedge}(\epsilon\overline{\epsilon}^{(4)})_2\nonumber\\
\delta(\epsilon\overline{\epsilon}^{(4)})_1&=&0.
\end{eqnarray}
Then, the reduction of the 1-form bilinear onto $M_4$ is also a KY 1-form
\begin{equation}
\nabla_{X_a}(\epsilon\overline{\epsilon}^{(4)})_1=\frac{1}{2}i_{X_a}d(\epsilon\overline{\epsilon}^{(4)})_1.
\end{equation}
Similarly, the 2-form bilinear equation (46) decomposes as
\begin{eqnarray}
\nabla_{X_a}(\epsilon\overline{\epsilon}^{(4)})_2&=&-\frac{i\lambda}{3}(\epsilon\overline{\epsilon}^{(4)})_1\underset{1}{\wedge}i_{X_a}z_4\nonumber\\
\nabla_{X_{\alpha}}(\epsilon\overline{\epsilon}^{(7)})_2&=&0.
\end{eqnarray}
From the first equality, we have
\begin{eqnarray}
d(\epsilon\overline{\epsilon}^{(4)})_2&=&i\lambda z_4\underset{1}{\wedge}(\epsilon\overline{\epsilon}^{(4)})_1\nonumber\\
\delta(\epsilon\overline{\epsilon}^{(4)})_2&=&0
\end{eqnarray}
and so $(\epsilon\overline{\epsilon}^{(4)})_2$ is a KY 2-form
\begin{eqnarray}
\nabla_{X_a}(\epsilon\overline{\epsilon}^{(4)})_2=\frac{1}{3}i_{X_a}d(\epsilon\overline{\epsilon}^{(4)})_2.
\end{eqnarray}
However, the second equality in (62) implies that $(\epsilon\overline{\epsilon}^{(7)})_2$ is a parallel form and hence must be constructed from the parallel spinor $\epsilon_7$. This implies from (22) that $\lambda$ must also vanish $\lambda=0$. Then, the bilinears on $M_4$ also correspond to parallel forms and $\epsilon_4$ is also a parallel spinor. So, the choice of inner product forces the flux $F$ to vanish and the decompositions of 5-form and 6-form bilinear equations also imply this. Then, as a result, the first inner product choice allows only $\text{Mink}_4\times G_2$ solutions.

ii) $M_4 : \mathbb{C}^*\text{-sym }\xi$ and $M_7 : \mathbb{R}\text{-sym }\xi\eta$;

In this case, all bilinear forms on $M_7$ which appear in the bilinear form equations are automatically zero as can be seen from Table I. So, the seven-dimensional parts of the decompositions are trivial and this does not give a restriction on $\mu$. The inner product choice for $M_4$ is the same as for the first case and hence the four-dimensional parts of the bilinears correspond to KY forms
\begin{eqnarray}
\nabla_{X_a}(\epsilon\overline{\epsilon}^{(4)})_1&=&\frac{1}{2}i_{X_a}d(\epsilon\overline{\epsilon}^{(4)})_1\nonumber\\
\nabla_{X_a}(\epsilon\overline{\epsilon}^{(4)})_2&=&\frac{1}{3}i_{X_a}d(\epsilon\overline{\epsilon}^{(4)})_2
\end{eqnarray}
and moreover they correspond to special KY forms. By direct computation, one can see that
\begin{eqnarray}
\nabla_{X_a}d(\epsilon\overline{\epsilon}^{(4)})_1&=&\frac{2}{9}\lambda^2e_a\wedge(\epsilon\overline{\epsilon}^{(4)})_1\nonumber\\
\nabla_{X_a}d(\epsilon\overline{\epsilon}^{(4)})_2&=&\frac{1}{3}\lambda^2e_a\wedge(\epsilon\overline{\epsilon}^{(4)})_2
\end{eqnarray}
and this implies that $\epsilon_4$ must correspond to a geometric Killing spinor. This also does not put a restriction on $\lambda$ and hence all types of solutions for this inner product choice is possible; $AdS_4\times \text{weak }G_2$, $AdS_4\times S^7$ and $\text{Mink}_4\times G_2$.

iii) $M_4 : \mathbb{C}\text{-skew }\xi\eta$ and $M_7 : \mathbb{R}\text{-skew }\xi$;

The choice of spinor inner product on $M_7$ is same as in the first case. So, this choice also implies that $\lambda=0$ and $\mu=0$ and  hence all the bilinears on $M_4$ and $M_7$ correspond to parallel forms. The only solution is $\text{Mink}_4\times G_2$.

iv) $M_4 : \mathbb{C}\text{-skew }\xi\eta$ and $M_7 : \mathbb{R}\text{-sym }\xi\eta$;

The choice of spinor inner product on $M_7$ is same as in the second case. So, there is no restriction on $\lambda$ and $\mu$ and bilinear forms on $M_4$ correspond to special KY forms. However, the choice of inner product on $M_4$ does not determine the real or pure imaginary character of the bilinear 1- and 2-forms and hence only if $\lambda$ can be chosen as real, we can have the solutions $AdS_4\times\text{weak }G_2$ and $AdS_4\times S^7$. $\text{Mink}_4\times G_2$ solution already exists since $\lambda=0=\mu$ for it.

In summary, the relation between spinor inner product choices and $M_4\times M_7$ solutions is as given in Table II.

\begin{table}[h]
\centering
\begin{tabular}{c|c}

$M_{11} : \mathbb{R}\text{-skew }\xi\eta$ & $\text{solutions}$\\ \hline
$M_4 : \mathbb{C}^*\text{-sym }\xi$ & $\text{Mink}_4\times G_2$\\
$M_7 : \mathbb{R}\text{-skew }\xi$ & $$\\ \hline
$M_4 : \mathbb{C}^*\text{-sym }\xi$ & $\text{Mink}_4\times G_2$\\
$M_7 : \mathbb{R}\text{-sym }\xi\eta$ & $AdS_4\times S^7\,,\,AdS_4\times\text{weak }G_2$\\ \hline
$M_4 : \mathbb{C}\text{-skew }\xi\eta$ & $\text{Mink}_4\times G_2$\\
$M_7 : \mathbb{R}\text{-skew }\xi$ & $$\\ \hline
$M_4 : \mathbb{C}\text{-skew }\xi\eta$ & $\text{Mink}_4\times G_2$\\
$M_7 : \mathbb{R}\text{-sym }\xi\eta$ & $AdS_4\times S^7\,,\,AdS_4\times\text{weak }G_2\,\,\text{(if }\lambda\text{ is real)}$\\

\end{tabular}
\caption{The relation between the choice of spinor inner products and $M_4\times M_7$ solutions for $\mathbb{R}$-skew $\xi\eta$ inner product on $M_{11}$.}
\end{table}

Note that $AdS$ solutions can exist only for the inner product choices for which the supergravity Killing forms decompose into special KY forms on product manifolds.

\subsubsection{$\mathbb{R}$-sym $\xi$ inner product}

In the second case, we choose the spinor inner product on the eleven-dimensional Lorentzian manifold $M_{11}$ as $\mathbb{R}$-sym with $\xi$ involution and consider the decomposition of bilinear forms in that case. $p$-form bilinear equation is the same as in (34)
\begin{equation}
\nabla_X(\epsilon\overline{\epsilon})_p=-\frac{1}{24}\left((\widetilde{X}.F-3F.\widetilde{X}).\epsilon\overline{\epsilon}\right)_p-\frac{1}{24}\left(\epsilon\overline{(\widetilde{X}.F-3F.\widetilde{X}).\epsilon}\right)_p.
\end{equation}
However, in this case the involution is $\xi$ and for a Clifford form $\omega$ and a spinor $\psi$, we have $\overline{\omega.\psi}=\psi^{\xi}.\omega^{\xi}=\overline{\psi}.\omega^{\xi}$. So, we can write
\begin{equation}
\overline{(\widetilde{X}.F-3F.\widetilde{X}).\epsilon}=\overline{\epsilon}.(F.\widetilde{X}-3\widetilde{X}.F)
\end{equation}
where we have used $F^{\xi}=F$ and ${\widetilde{X}}^{\xi}=\widetilde{X}$. By adding and subtracting the term $\frac{1}{6}\left(\epsilon\overline{\epsilon}.(F.\widetilde{X}-\widetilde{X}.F)\right)_p$ to (67), we find
\begin{equation}
\nabla_X(\epsilon\overline{\epsilon})_p=-\frac{1}{24}\left([(\widetilde{X}.F-3F.\widetilde{X}),\epsilon\overline{\epsilon}]_{+Cl}\right)_p-\frac{1}{6}\left(\epsilon\overline{\epsilon}.[F,\widetilde{X}]_{Cl}\right)_p
\end{equation}
where $[\,,\,]_{+Cl}$ denotes the Clifford anticommutator which is defined in (B11) and (B12). In terms of wedge product and interior derivative, the supergravity Killing form equation can also be written from (40) as
\begin{equation}
\nabla_X(\epsilon\overline{\epsilon})_p=\frac{1}{12}\left([\widetilde{X}\wedge F, \epsilon\overline{\epsilon}]_{+Cl}\right)_p-\frac{1}{6}\left([i_XF, \epsilon\overline{\epsilon}]_{+Cl}\right)_p+\frac{1}{3}\left(\epsilon\overline{\epsilon}.i_XF\right)_p.
\end{equation}
The nonzero bilinear forms for $\mathbb{R}$-sym $\xi$ inner product are 0-, 1-, 4-, 5-, 8- and 9-forms and the spinor bilinear of the supergravity Killing spinor $\epsilon$ corresponds to
\begin{equation}
\epsilon\overline{\epsilon}=(\epsilon\overline{\epsilon})_0+(\epsilon\overline{\epsilon})_1+(\epsilon\overline{\epsilon})_4+(\epsilon\overline{\epsilon})_5+(\epsilon\overline{\epsilon})_8+(\epsilon\overline{\epsilon})_9.
\end{equation}
From (70), we can find the bilinear form equations for different degrees. For $p=0$, we have
\begin{equation}
\nabla_{X_A}(\epsilon\overline{\epsilon})_0=\frac{1}{144}F\underset{4}{\wedge}i_{X_A}(\epsilon\overline{\epsilon})_5
\end{equation} 
and
\begin{eqnarray}
d(\epsilon\overline{\epsilon})_0&=&\frac{1}{144}F\underset{4}{\wedge}(\epsilon\overline{\epsilon})_5\nonumber\\
\delta(\epsilon\overline{\epsilon})_0&=&0.
\end{eqnarray}
Then $(\epsilon\overline{\epsilon})_0$ satisfies $\nabla_{X_A}(\epsilon\overline{\epsilon})_0=i_{X_A}d(\epsilon\overline{\epsilon})_0$ and hence is a KY 0-form. For $p=1$, the bilinear form equation corresponds to
\begin{equation}
\nabla_{X_A}(\epsilon\overline{\epsilon})_1=\frac{1}{144}(e_A\wedge F)\underset{4}{\wedge}(\epsilon\overline{\epsilon})_4+\frac{1}{18}(\epsilon\overline{\epsilon})_4\underset{3}{\wedge}i_{X_A}F
\end{equation}
and
\begin{eqnarray}
d(\epsilon\overline{\epsilon})_1&=&-\frac{1}{36}F\underset{3}{\wedge}(\epsilon\overline{\epsilon})_4\nonumber\\
\delta(\epsilon\overline{\epsilon})_1&=&\frac{1}{144}F\underset{4}{\wedge}(\epsilon\overline{\epsilon})_4.
\end{eqnarray}
For $p=4$, (70) gives
\begin{eqnarray}
\nabla_{X_A}(\epsilon\overline{\epsilon})_4&=&\frac{1}{6}(e_A\wedge F)\underset{1}{\wedge}(\epsilon\overline{\epsilon})_1-\frac{1}{36}(e_A\wedge F)\underset{3}{\wedge}(\epsilon\overline{\epsilon})_5+\frac{1}{144}F\underset{4}{\wedge}i_{X_A}(\epsilon\overline{\epsilon})_9\nonumber\\
&&+\frac{1}{3}(\epsilon\overline{\epsilon})_1\wedge i_{X_A}F+\frac{1}{6}i_{X_A}F\underset{2}{\wedge}(\epsilon\overline{\epsilon})_5
\end{eqnarray}
and
\begin{eqnarray}
d(\epsilon\overline{\epsilon})_4&=&-\frac{7}{6}F\wedge(\epsilon\overline{\epsilon})_1+\frac{1}{12}F\underset{2}{\wedge}(\epsilon\overline{\epsilon})_5+\frac{5}{144}F\underset{4}{\wedge}(\epsilon\overline{\epsilon})_9\nonumber\\
\delta(\epsilon\overline{\epsilon})_4&=&\frac{5}{6}F\underset{1}{\wedge}(\epsilon\overline{\epsilon})_1-\frac{1}{18}F\underset{3}{\wedge}(\epsilon\overline{\epsilon})_5.
\end{eqnarray}
For $p=5$, we have
\begin{eqnarray}
\nabla_{X_A}(\epsilon\overline{\epsilon})_5&=&\frac{1}{6}e_A\wedge F\wedge(\epsilon\overline{\epsilon})_0-\frac{1}{12}(e_A\wedge F)\underset{2}{\wedge}(\epsilon\overline{\epsilon})_4+\frac{1}{144}(e_A\wedge F)\underset{4}{\wedge}(\epsilon\overline{\epsilon})_8\nonumber\\
&&-\frac{1}{3}(\epsilon\overline{\epsilon})_4\underset{1}{\wedge}i_{X_A}F+\frac{1}{18}(\epsilon\overline{\epsilon})_8\underset{3}{\wedge}i_{X_A}F
\end{eqnarray}
and
\begin{eqnarray}
d(\epsilon\overline{\epsilon})_5&=&\frac{1}{2}F\underset{1}{\wedge}(\epsilon\overline{\epsilon})_4+\frac{1}{12}F\underset{3}{\wedge}(\epsilon\overline{\epsilon})_8\nonumber\\
\delta(\epsilon\overline{\epsilon})_5&=&-\frac{7}{6}F\wedge(\epsilon\overline{\epsilon})_0+\frac{13}{12}F\underset{2}{\wedge}(\epsilon\overline{\epsilon})_4-\frac{19}{144}F\underset{4}{\wedge}(\epsilon\overline{\epsilon})_8.
\end{eqnarray}
Similarly, for $p=8$, (70) gives
\begin{equation}
\nabla_{X_A}(\epsilon\overline{\epsilon})_8=\frac{1}{6}(e_A\wedge F)\underset{1}{\wedge}(\epsilon\overline{\epsilon})_5-\frac{1}{36}(e_A\wedge F)\underset{3}{\wedge}(\epsilon\overline{\epsilon})_9-\frac{1}{3}i_{X_A}F\wedge(\epsilon\overline{\epsilon})_5+\frac{1}{6}i_{X_A}F\underset{2}{\wedge}(\epsilon\overline{\epsilon})_9
\end{equation}
and
\begin{eqnarray}
d(\epsilon\overline{\epsilon})_8&=&-\frac{1}{2}F\wedge(\epsilon\overline{\epsilon})_5-\frac{1}{4}F\underset{2}{\wedge}(\epsilon\overline{\epsilon})_9\nonumber\\
\delta(\epsilon\overline{\epsilon})_8&=&\frac{1}{6}F\underset{1}{\wedge}(\epsilon\overline{\epsilon})_5+\frac{5}{36}F\underset{3}{\wedge}(\epsilon\overline{\epsilon})_9.
\end{eqnarray}
For $p=9$, we have
\begin{equation}
\nabla_{X_A}(\epsilon\overline{\epsilon})_9=\frac{1}{6}e_A\wedge F\wedge(\epsilon\overline{\epsilon})_4-\frac{1}{12}(e_A\wedge F)\underset{2}{\wedge}(\epsilon\overline{\epsilon})_8-\frac{1}{3}(\epsilon\overline{\epsilon})_8\underset{1}{\wedge}i_{X_A}F
\end{equation}
and
\begin{eqnarray}
d(\epsilon\overline{\epsilon})_9&=&-\frac{1}{6}F\underset{1}{\wedge}(\epsilon\overline{\epsilon})_8\nonumber\\
\delta(\epsilon\overline{\epsilon})_9&=&-\frac{1}{2}F\wedge(\epsilon\overline{\epsilon})_4+\frac{3}{4}F\underset{2}{\wedge}(\epsilon\overline{\epsilon})_8.
\end{eqnarray}
So, only the 0-form bilinear correspond to a KY form and other higher degree bilinears satisfy different types of equations.

Now, we can decompose the bilinear form equations onto product manifolds $M_4$ and $M_7$. We have the following bilinear forms on product manifolds
\begin{eqnarray}
\epsilon\overline{\epsilon}^{(4)}&=&(\epsilon\overline{\epsilon}^{(4)})_0+(\epsilon\overline{\epsilon}^{(4)})_1+(\epsilon\overline{\epsilon}^{(4)})_4\nonumber\\
\epsilon\overline{\epsilon}^{(7)}&=&(\epsilon\overline{\epsilon}^{(7)})_0+(\epsilon\overline{\epsilon}^{(7)})_1+(\epsilon\overline{\epsilon}^{(7)})_4+(\epsilon\overline{\epsilon}^{(7)})_5
\end{eqnarray}
and from Table XVII in Appendix A, the properties of bilinear forms depending on the choice of the spinor inner product are given in Table III.

\begin{table}[h]
\centering
\begin{tabular}{c c|c c c c}

$$ & $\text{inner product}$ & $0$ & $1$ & $4$ & $5$\\ \hline
$i)$ & $M_4 : \mathbb{C}^*\text{-sym }\xi$ & $R$ & $R$ & $R$ & $$\\
$$ & $M_7 : \mathbb{R}\text{-skew }\xi$ & $\times$ & $\times$ & $\times$ & $\times$\\ \hline
$ii)$ & $M_4 : \mathbb{C}^*\text{-sym }\xi$ & $R$ & $R$ & $R$ & $$\\
$$ & $M_7 : \mathbb{R}\text{-sym }\xi\eta$ & $\checkmark$ & $\times$ & $\checkmark$ & $\times$\\ \hline
$iii)$ & $M_4 : \mathbb{C}\text{-skew }\xi\eta$ & $\times$ & $\checkmark$ & $\times$ & $$\\
$$ & $M_7 : \mathbb{R}\text{-skew }\xi$ & $\times$ & $\times$ & $\times$ & $\times$\\ \hline
$iv)$ & $M_4 : \mathbb{C}\text{-skew }\xi\eta$ & $\times$ & $\checkmark$ & $\times$ & $$\\
$$ & $M_7 : \mathbb{R}\text{-sym }\xi\eta$ & $\checkmark$ & $\times$ & $\checkmark$ & $\times$\\

\end{tabular}
\caption{Properties of nonzero bilinears for different spinor inner product choices on Lorentzian $M_4$ and Riemannian $M_7$ for $\mathbb{R}$-sym $\xi$ inner product on $M_{11}$.}
\end{table}

Then, we can consider four different cases in the decomposition.

i) $M_4 : \mathbb{C}^*\text{-sym }\xi$ and $M_7 : \mathbb{R}\text{-skew }\xi$;

For this inner product choice, the bilinear form equations on $M_4$ correspond to
\begin{eqnarray}
\nabla_{X_a}(\epsilon\overline{\epsilon}^{(4)})_0&=&0\nonumber\\
\nabla_{X_a}(\epsilon\overline{\epsilon}^{(4)})_1&=&\frac{i\lambda}{36}z_4\underset{3}{\wedge}i_{X_a}(\epsilon\overline{\epsilon}^{(4)})_4+\frac{i\lambda}{144}e_a\wedge(z_4\underset{4}{\wedge}(\epsilon\overline{\epsilon}^{(4)})_4)+\frac{i\lambda}{18}(\epsilon\overline{\epsilon}^{(4)})_4\underset{3}{\wedge}i_{X_a}z_4\\
\nabla_{X_a}(\epsilon\overline{\epsilon}^{(4)})_4&=&\frac{i\lambda}{3}(\epsilon\overline{\epsilon}^{(4)})_1\wedge i_{X_a}z_4.\nonumber
\end{eqnarray}
So, $(\epsilon\overline{\epsilon}^{(4)})_0$ is constant and we can write the exterior and coderivatives of bilinear forms as
\begin{eqnarray}
d(\epsilon\overline{\epsilon}^{(4)})_1&=&0\nonumber\\
\delta(\epsilon\overline{\epsilon}^{(4)})_1&=&\frac{i\lambda}{18}z_4\underset{1}{\wedge}(\epsilon\overline{\epsilon}^{(4)})_4
\end{eqnarray}
and
\begin{eqnarray}
d(\epsilon\overline{\epsilon}^{(4)})_4&=&0\nonumber\\
\delta(\epsilon\overline{\epsilon}^{(4)})_4&=&-\frac{i\lambda}{3}(\epsilon\overline{\epsilon}^{(4)})_1\underset{1}{\wedge}z_4.
\end{eqnarray}
Then, by comparing (85) with (86) and (87), one can see that they satisfy the CCKY equation
\begin{eqnarray}
\nabla_{X_a}(\epsilon\overline{\epsilon}^{(4)})_1&=&-\frac{1}{4}e_a\wedge\delta(\epsilon\overline{\epsilon}^{(4)})_1\nonumber\\
\nabla_{X_a}(\epsilon\overline{\epsilon}^{(4)})_4&=&-e_a\wedge\delta(\epsilon\overline{\epsilon}^{(4)})_4.
\end{eqnarray}
Moreover, they correspond to special CCKY forms
\begin{eqnarray}
\nabla_{X_a}\delta(\epsilon\overline{\epsilon}^{(4)})_1&=&-\frac{4\lambda^2}{9}i_{X_a}(\epsilon\overline{\epsilon}^{(4)})_1\nonumber\\
\nabla_{X_a}\delta(\epsilon\overline{\epsilon}^{(4)})_4&=&-\frac{\lambda^2}{9}i_{X_a}(\epsilon\overline{\epsilon}^{(4)})_4.
\end{eqnarray}
So, $\epsilon_4$ is a geometric Killing spinor generating the supergravity Killing forms which correspond to special CCKY forms. All of the bilinear form equations on $M_7$ are trivial and hence we have all types of solutions for this inner product choice. Namely, $AdS_4\times \text{weak }G_2$, $AdS_4\times S^7$ and $\text{Mink}_4\times G_2$.

ii) $M_4 : \mathbb{C}^*\text{-sym }\xi$ and $M_7 : \mathbb{R}\text{-sym }\xi\eta$;

In this case, the situation for $M_4$ is the same as in the previous case and hence $(\epsilon\overline{\epsilon}^{(4)})_1$ and $(\epsilon\overline{\epsilon}^{(4)})_4$ are special CCKY forms. For $M_7$, we have the following equalities
\begin{eqnarray}
\nabla_{X_{\alpha}}(\epsilon\overline{\epsilon}^{(7)})_0&=&0\nonumber\\
0&=&\frac{\mu}{36}\phi\underset{3}{\wedge}i_{X_{\alpha}}(\epsilon\overline{\epsilon}^{(7)})_4+\frac{\mu}{18}(\epsilon\overline{\epsilon}^{(7)})_4\underset{3}{\wedge}i_{X_{\alpha}}\phi+\frac{\mu}{144}e_{\alpha}\wedge(\phi\underset{4}{\wedge}(\epsilon\overline{\epsilon}^{(7)})_4)\nonumber\\
\nabla_{X_{\alpha}}(\epsilon\overline{\epsilon}^{(7)})_4&=&0\\
0&=&\frac{\mu}{6}e_{\alpha}\wedge\phi\wedge(\epsilon\overline{\epsilon}^{(7)})_0-\frac{\mu}{12}(e_{\alpha}\wedge\phi)\underset{2}{\wedge}(\epsilon\overline{\epsilon}^{(7)})_4-\frac{\mu}{3}(\epsilon\overline{\epsilon}^{(7)})_4\underset{1}{\wedge}i_{X_{\alpha}}\phi.\nonumber
\end{eqnarray}
So, we have $\mu=0$ and 0- and 4-forms are parallel. Then, we have the solutions $AdS_4\times\text{weak }G_2$, $AdS_4\times S^7$ for $\lambda\neq0$ and $\mu=0$. For $\lambda=\mu=0$, we have $\text{Mink}_4\times G_2$.

iii) $M_4 : \mathbb{C}\text{-skew }\xi\eta$ and $M_7 : \mathbb{R}\text{-skew }\xi$;

This case gives 
\begin{eqnarray}
\nabla_{X_a}(\epsilon\overline{\epsilon}^{(4)})_1&=&0\nonumber\\
0&=&\frac{i\lambda}{3}(\epsilon\overline{\epsilon}^{(4)})_1\wedge i_{X_a}z_4.
\end{eqnarray}
on $M_4$ and we have $\lambda=0$. The seven-dimensional equations on $M_7$ are all trivial and the only solution is $\text{Mink}_4\times G_2$.

iv) $M_4 : \mathbb{C}\text{-skew }\xi\eta$ and $M_7 : \mathbb{R}\text{-sym }\xi\eta$;

The case for $M_4$ is the same as the previous case and for $M_7$ it is the same with case (ii). So, both $\lambda$ and $\mu$ vanishes and we have $\text{Mink}_4\times G_2$ solution.

In summary, for the inner product choice of $\mathbb{R}$-sym $\xi$ on $M_{11}$, the solutions that appear for different types of inner product choices on $M_4$ and $M_7$ can be given as in Table IV.

\begin{table}
\centering
\begin{tabular}{c|c}

$M_{11} : \mathbb{R}\text{-sym }\xi$ & $\text{solutions}$\\ \hline
$M_4 : \mathbb{C}^*\text{-sym }\xi$ & $\text{Mink}_4\times G_2$\\
$M_7 : \mathbb{R}\text{-skew }\xi$ & $AdS_4\times S^7\,,\,AdS_4\times\text{weak }G_2$\\ \hline
$M_4 : \mathbb{C}^*\text{-sym }\xi$ & $\text{Mink}_4\times G_2$\\
$M_7 : \mathbb{R}\text{-sym }\xi\eta$ & $AdS_4\times S^7\,,\,AdS_4\times\text{weak }G_2$\\ \hline
$M_4 : \mathbb{C}\text{-skew }\xi\eta$ & $\text{Mink}_4\times G_2$\\
$M_7 : \mathbb{R}\text{-skew }\xi$ & $$\\ \hline
$M_4 : \mathbb{C}\text{-skew }\xi\eta$ & $\text{Mink}_4\times G_2$\\
$M_7 : \mathbb{R}\text{-sym }\xi\eta$ & $$\\

\end{tabular}
\caption{The relation between the choice of spinor inner products and $M_4\times M_7$ solutions for $\mathbb{R}$-sym $\xi$ inner product on $M_{11}$.}
\end{table}

Note that, when the $AdS$ solutions exist for the relevant choices of spinor inner products, the supergravity Killing forms decompose into special CCKY forms on product manifolds.

\section{$M_7\times M_4$ Type Backgrounds}

In another case, we consider the product structure of the eleven-dimensional background $M_{11}$ as $M_{11}=M_7\times M_4$ where $M_7$ is a Lorentzian spin 7-manifold and $M_4$ is a Riemannian spin 4-manifold. The eleven-dimensional indices will split into $A=\{a,\alpha\}$ with $a=0,1,2,...,6$ and $\alpha=7,8,9,10$. The Clifford algebra basis is decomposed as
\begin{equation}
e^A=\{1_7\otimes e^{\alpha}, e^a\otimes z_4\}
\end{equation}
where $1_7$ is the identity on $M_7$, $e^a$ are the Clifford algebra basis on $M_7$, $z_4$ is the volume form on $M_4$ and $e^{\alpha}$ are the Clifford algebra basis on $M_4$. The flux 4-form is decomposed as
\begin{equation}
F=\{\lambda\phi,\mu z_4\}
\end{equation}
where $\lambda$ and $\mu$ are constants and $\phi$ is a 4-form on $M_7$. Similarly, the supersymmetry parameter can be written as
\begin{equation}
\epsilon=\epsilon_7\otimes\epsilon_4.
\end{equation}
By decomposing the field equations similar to the case in Section III, the Maxwell-like field equations give
\begin{eqnarray}
d*_7\phi&=&\mu\phi\nonumber\\
d\phi&=&0.
\end{eqnarray}
This means that $\phi$ is a special CCKY 4-form on $M_7$ and must be generated from a geometric Killing spinor. Since the volume form $z_4$ is also a special KY 4-form, the flux form $F$ is generated by KY and CCKY forms and hence by geometric Killing spinors for $\lambda\neq  0$ and $\mu\neq 0$.

The decomposition of Einstein field equations will give the following equalities on $M_4$ and $M_7$ respectively
\begin{eqnarray}
i_{X_b}P_a&=&\frac{\lambda^2}{2}\bigg(g_3(i_{X_a}\phi,i_{X_b}\phi)-\frac{1}{3}g_4(\phi,\phi)g_{ab}\bigg)-\frac{\mu^2}{6}g_{ab}\\
i_{X_{\beta}}P_{\alpha}&=&\frac{1}{3}\bigg(\mu^2-\frac{\lambda^2}{2}g_4(\phi,\phi)\bigg)g_{\alpha\beta}.
\end{eqnarray}
This means that for $\lambda=\mu=0$, both $M_7$ and $M_4$ are Ricci-flat manifolds and for the special case of $\lambda=0$ and $\mu\neq 0$, $M_7$ is a negative curvature and $M_4$ is a positive curvature Einstein manifolds.

The decomposition of the supergravity Killing spinor equation into product manifolds can be found as follows
\begin{eqnarray}
\nabla_{X^a}\epsilon_7\otimes\epsilon_4+\epsilon_7\otimes\nabla_{X^{\alpha}}\epsilon_4&=&\frac{1}{12}\lambda\phi.\epsilon_7\otimes e^{\alpha}.\epsilon_4+\frac{1}{12}\mu e^a.\epsilon_7\otimes\epsilon_4\nonumber\\
&&\mp\frac{1}{6}\epsilon_7\otimes\mu e^{\alpha}.\epsilon_4\mp\frac{1}{24}\lambda(e^a.\phi-3\phi.e^a).\epsilon_7\otimes\epsilon_4
\end{eqnarray}
where we have used that $z_4.e^{\alpha}=-e^{\alpha}.z_4$ and $z_4^2=1$ for the Riemannian manifold $M_4$, and we also require $z_4.\epsilon_4=\pm\epsilon_4$. Then, for the fluxless case $\lambda=\mu=0$, we have two equations on product manifolds
\begin{eqnarray}
\nabla_{X^a}\epsilon_7&=&0\nonumber\\
\nabla_{X^{\alpha}}\epsilon_4&=&0
\end{eqnarray}
and this means that $\epsilon_7$ and $\epsilon_4$ are parallel spinors on $M_7$ and $M_4$, respectively and both $M_7$ and $M_4$ are Ricci-flat manifolds. 4-dimensional Riemannian manifolds admitting parallel spinors correspond to Calabi-Yau manifolds with $SU(2)$ holonomy (and also hyperk\"{a}hler manifolds with $Sp(1)$ holonomy but they are equivalent to Calabi-Yau manifolds with $SU(2)$ holonomy). 7-dimensional Ricci-flat Lorentzian manifolds admitting parallel spinors can be Minkowski spacetimes. So, fluxless case corresponds to $\text{Mink}_7\times CY_2$. For the existence of only internal flux $\lambda=0$ and $\mu\neq 0$, we have the following equations
\begin{eqnarray}
\nabla_{X^a}\epsilon_7&=&\frac{\mu}{12}e^a.\epsilon_7\nonumber\\
\nabla_{X^{\alpha}}\epsilon_4&=&\mp\frac{\mu}{6}e^{\alpha}.\epsilon_4
\end{eqnarray}
and hence $\epsilon_7$ and $\epsilon_4$ correspond to geometric Killing spinors on $M_7$ and $M_4$, respectively. So, $M_7$ and $M_4$ are Einstein manifolds. The only four dimensional Riemannian manifold admitting geometric Killing spinors is the four-sphere $S^4$ and the solution in this case corresponds to $AdS_7\times S^4$. If both internal and external fluxes are non-zero $\lambda\neq 0$ and $\mu\neq 0$ and $\phi$ satisfies $\phi.\epsilon_7=\pm\frac{1}{2}\epsilon_7$, then the supergravity Killing spinor equation decomposes into
\begin{eqnarray}
\nabla_{X^a}\epsilon_7&=&\frac{1}{12}\left(\mu-\frac{\lambda}{4}\right)e^a.\epsilon_7\pm\frac{\lambda}{8}\phi.e^a.\epsilon_7\\
\nabla_{X^{\alpha}}\epsilon_4&=&\pm\frac{1}{6}\left(\frac{\lambda}{4}-\mu\right)e^{\alpha}.\epsilon_4.
\end{eqnarray}
By doing similar calculations as in Section III, one can find that if $\phi$ satisfies the condition
\begin{equation}
(\mathcal{L}_{X^a}*_7\phi).\epsilon_7=\mu e^a.\epsilon_7
\end{equation}
then (101) transforms into a geometric Killing spinor equation and both $\epsilon_7$ and $\epsilon_4$ are geometric Killing spinors. However, this case does not give a new solution and also corresponds to $AdS_7\times S^4$ solution. For the case of $\lambda\neq 0$ and $\mu=0$, the field equations and Killing spinor equations give an inconsistency and hence this case does not correspond to a solution.

The decomposition of bilinear forms of supergravity Killing spinors have to be investigated separately for different choices of spinor inner products. For the choice of spinor inner product $\mathbb{R}$-skew $\xi\eta$ on $M_{11}$, the supergravity Killing forms $(\epsilon\overline{\epsilon})_p=\{(\epsilon\overline{\epsilon}^{(7)})_p,(\epsilon\overline{\epsilon}^{(4)})_p\}$ satisfy (41) and non-zero bilinear forms correspond to
\begin{eqnarray}
\epsilon\overline{\epsilon}^{(7)}&=&(\epsilon\overline{\epsilon}^{(7)})_1+(\epsilon\overline{\epsilon}^{(7)})_2+(\epsilon\overline{\epsilon}^{(7)})_5+(\epsilon\overline{\epsilon}^{(7)})_6\nonumber\\
\epsilon\overline{\epsilon}^{(4)}&=&(\epsilon\overline{\epsilon}^{(4)})_1+(\epsilon\overline{\epsilon}^{(4)})_2\nonumber
\end{eqnarray}
on $M_7$ and $M_4$, respectively. $M_7$ is a Lorentzian 7-manifold, so the spinor space is $\mathbb{H}^4$ and the spinors are symplectic Majorana spinors. $M_4$ is a Riemannian 4-manifold, so the spinor space is $\mathbb{H}\oplus\mathbb{H}$ and the spinors are symplectic Majorana-Weyl spinors. From Table XVII in Appendix A, we have the bilinears for the chosen inner products given in Table V.

\begin{table}
\centering
\begin{tabular}{c c|c c c c}

$$ & $\text{inner product}$ & $1$ & $2$ & $5$ & $6$\\ \hline
$i)$ & $M_7 : \mathbb{H}^{-}\text{-sym }\xi$ & $R$ & $V$ & $R$ & $V$\\
$$ & $M_4 : \mathbb{H}\text{-swap }\xi$ & $\checkmark$ & $\times$ & $$ & $$\\ \hline
$ii)$ & $M_7 : \mathbb{H}^{-}\text{-sym }\xi$ & $R$ & $V$ & $R$ & $V$\\
$$ & $M_4 : \mathbb{H}^{-}\text{-sym}\oplus\mathbb{H}^{-}\text{-sym }\xi\eta$ & $V$ & $V$ & $$ & $$\\ \hline
$iii)$ & $M_7 : \mathbb{H}\,\widehat{\,}\text{-sym }\xi\eta$ & $(-)$ & $(-)$ & $(-)$ & $(-)$\\
$$ & $M_4 : \mathbb{H}\text{-swap }\xi$ & $\checkmark$ & $\times$ & $$ & $$\\ \hline
$iv)$ & $M_7 : \mathbb{H}\,\widehat{\,}\text{-sym }\xi\eta$ & $(-)$ & $(-)$ & $(-)$ & $(-)$\\
$$ & $M_4 : \mathbb{H}^{-}\text{-sym}\oplus\mathbb{H}^{-}\text{-sym }\xi\eta$ & $V$ & $V$ & $$ & $$\\

\end{tabular}
\caption{Properties of nonzero bilinears for different spinor inner product choices on Lorentzian $M_7$ and Riemannian $M_4$ for $\mathbb{R}$-skew $\xi\eta$ inner product on $M_{11}$.}
\end{table}

We consider four different inner product choices in the decomposition of supergravity Killing forms.

i) $M_7 : \mathbb{H}^{-}\text{-sym }\xi$ and $M_4 : \mathbb{H}\text{-swap }\xi$;

In that case, the bilinear form equations for $(\epsilon\overline{\epsilon})_1$, $(\epsilon\overline{\epsilon})_2$, $(\epsilon\overline{\epsilon})_5$, $(\epsilon\overline{\epsilon})_6$, $(\epsilon\overline{\epsilon})_9$ and $(\epsilon\overline{\epsilon})_{10}$ on $M_7$ corresponds to
\begin{eqnarray}
\nabla_{X_a}(\epsilon\overline{\epsilon}^{(7)})_1&=&\frac{\lambda}{144}\phi\underset{4}{\wedge}i_{X_a}(\epsilon\overline{\epsilon}^{(7)})_6-\frac{\lambda}{6}i_{X_a}\phi\underset{2}{\wedge}(\epsilon\overline{\epsilon}^{(7)})_2\nonumber\\
\nabla_{X_a}(\epsilon\overline{\epsilon}^{(7)})_2&=&\frac{\lambda}{36}\phi\underset{3}{\wedge}i_{X_a}(\epsilon\overline{\epsilon}^{(7)})_5+\frac{\lambda}{144}e_a\wedge(\phi\underset{4}{\wedge}(\epsilon\overline{\epsilon}^{(7)})_5)\nonumber\\
&&-\frac{\lambda}{3}(\epsilon\overline{\epsilon}^{(7)})_1\underset{1}{\wedge}i_{X_a}\phi+\frac{\lambda}{18}(\epsilon\overline{\epsilon}^{(7)})_5\underset{3}{\wedge}i_{X_a}\phi\nonumber\\
\nabla_{X_a}(\epsilon\overline{\epsilon}^{(7)})_5&=&\frac{\lambda}{6}(e_a\wedge\phi)\underset{1}{\wedge}(\epsilon\overline{\epsilon}^{(7)})_2-\frac{\lambda}{3}i_{X_a}\phi\wedge(\epsilon\overline{\epsilon}^{(7)})_2\nonumber\\
&&+\frac{\lambda}{6}i_{X_a}\phi\underset{2}{\wedge}(\epsilon\overline{\epsilon}^{(7)})_6-\frac{\lambda}{36}(e_a\wedge\phi)\underset{3}{\wedge}(\epsilon\overline{\epsilon}^{(7)})_6\\
\nabla_{X_a}(\epsilon\overline{\epsilon}^{(7)})_6&=&-\frac{\lambda}{3}(\epsilon\overline{\epsilon}^{(7)})_5\underset{1}{\wedge}i_{X_a}\phi+\frac{\lambda}{6}(e_a\wedge\phi)\wedge(\epsilon\overline{\epsilon}^{(7)})_1-\frac{\lambda}{12}(e_a\wedge\phi)\underset{2}{\wedge}(\epsilon\overline{\epsilon}^{(7)})_5\nonumber\\
0&=&\frac{\lambda}{6}(e_a\wedge\phi)\underset{1}{\wedge}(\epsilon\overline{\epsilon}^{(7)})_6-\frac{\lambda}{3}i_{X_a}\phi
\wedge(\epsilon\overline{\epsilon}^{(7)})_6\nonumber\\
0&=&\frac{\lambda}{6}(e_a\wedge\phi)\wedge(\epsilon\overline{\epsilon}^{(7)})_5\nonumber
\end{eqnarray}
and on $M_4$, we have
\begin{eqnarray}
\nabla_{X_{\alpha}}(\epsilon\overline{\epsilon}^{(4)})_1&=&0\nonumber\\
0&=&-\frac{\mu}{3}(\epsilon\overline{\epsilon}^{(4)})_1\underset{1}{\wedge}i_{X_{\alpha}}z_4.
\end{eqnarray}
As can be seen from the last two equalities of (104) and the second equality in (105), both $\lambda$ and $\mu$ have to vanish, $\lambda=\mu=0$. Then bilinear forms are parallel and they are constructed from parallel spinors $\epsilon_7$ and $\epsilon_4$. Hence, this case corresponds to the fluxless case and we have $\text{Mink}_7\times CY_2$ solution.

ii) $M_7 : \mathbb{H}^{-}\text{-sym }\xi$ and $M_4 : \mathbb{H}^{-}\text{-sym}\oplus\mathbb{H}^{-}\text{-sym }\xi\eta$;

For this choice, the bilinear form equations on $M_7$ are same as in case (i) since the inner product is same. The equations on $M_4$ are as follows
\begin{eqnarray}
\nabla_{X_{\alpha}}(\epsilon\overline{\epsilon}^{(4)})_1&=&-\frac{\mu}{6}i_{X_{\alpha}}z_4\underset{2}{\wedge}(\epsilon\overline{\epsilon}^{(4)})_2\\
\nabla_{X_{\alpha}}(\epsilon\overline{\epsilon}^{(4)})_2&=&-\frac{\mu}{3}(\epsilon\overline{\epsilon}^{(4)})_1\underset{1}{\wedge}i_{X_{\alpha}}z_4.
\end{eqnarray}
Hence, we have $\lambda=0$ and $\mu\neq 0$. Since $(\epsilon\overline{\epsilon}^{(4)})_1$ and $(\epsilon\overline{\epsilon}^{(4)})_2$ are pure vector quantities from Table V, $\mu$ corresponds to a real number. Moreover, from (106) and (107), we can write
\begin{eqnarray}
d(\epsilon\overline{\epsilon}^{(4)})_1&=&-\frac{\mu}{3}z_4\underset{2}{\wedge}(\epsilon\overline{\epsilon}^{(4)})_2\\
\delta(\epsilon\overline{\epsilon}^{(4)})_1&=&0\nonumber
\end{eqnarray}
and
\begin{eqnarray}
d(\epsilon\overline{\epsilon}^{(4)})_2&=&\mu(\epsilon\overline{\epsilon}^{(4)})_1\underset{1}{\wedge}z_4\\
\delta(\epsilon\overline{\epsilon}^{(4)})_2&=&0.
\end{eqnarray}
So, $(\epsilon\overline{\epsilon}^{(4)})_1$ and $(\epsilon\overline{\epsilon}^{(4)})_2$ correspond to KY forms
\begin{eqnarray}
\nabla_{X_{\alpha}}(\epsilon\overline{\epsilon}^{(4)})_1&=&\frac{1}{2}i_{X_{\alpha}}d(\epsilon\overline{\epsilon}^{(4)})_1\nonumber\\
\nabla_{X_{\alpha}}(\epsilon\overline{\epsilon}^{(4)})_2&=&\frac{1}{3}i_{X_{\alpha}}d(\epsilon\overline{\epsilon}^{(4)})_2
\end{eqnarray}
and in fact they are special KY forms
\begin{eqnarray}
\nabla_{X_{\alpha}}d(\epsilon\overline{\epsilon}^{(4)})_1&=&\frac{2\mu^2}{9}e_{\alpha}\wedge(\epsilon\overline{\epsilon}^{(4)})_1\nonumber\\
\nabla_{X_{\alpha}}d(\epsilon\overline{\epsilon}^{(4)})_2&=&\frac{\mu^2}{3}e_{\alpha}\wedge(\epsilon\overline{\epsilon}^{(4)})_2.
\end{eqnarray}
Hence, this inner product choice coresponds to $AdS_7\times S^4$ solution and the geometric Killing spinor $\epsilon_7$ generates the flux component $\phi$ which is a special CCKY form and the geometric Killing spinor $\epsilon_4$ generates the bilinear forms $(\epsilon\overline{\epsilon}^{(4)})_1$ and $(\epsilon\overline{\epsilon}^{(4)})_2$ which are special KY forms. If we choose $\mu=0$, then the solution reduces to $\text{Mink}_7\times CY_2$ case and the bilinear forms correspond to parallel forms.

iii) $M_7 : \mathbb{H}\,\widehat{\,}\text{-sym }\xi\eta$ and $M_4 : \mathbb{H}\text{-swap }\xi$;

Since all the bilinear forms on $M_7$ are also nonzero in this inner product choice, the equations satisfied by the bilinear forms are the same as in the previous inner product choices. Similarly, the equations on $M_4$ are the same with case (i) because of the same inner product choice and we have $\lambda=\mu=0$ in that case. So, $\epsilon_7$ and $\epsilon_4$ are parallel spinors and this choice corresponds to $\text{Mink}_7\times CY_2$.

iv) $M_7 : \mathbb{H}\,\widehat{\,}\text{-sym }\xi\eta$ and $M_4 : \mathbb{H}^{-}\text{-sym}\oplus\mathbb{H}^{-}\text{-sym }\xi\eta$;

The bilinear form equations in this choice are exactly the same as in case (ii) and we have $\lambda=0$ and $\mu\neq 0$. So, the geometric Killing spinor $\epsilon_7$ generates the flux component $\phi$ and the geometric Killing spinor $\epsilon_4$ generates the bilinear forms $(\epsilon\overline{\epsilon}^{(4)})_1$ and $(\epsilon\overline{\epsilon}^{(4)})_2$ which are special KY forms. So, this case corresponds to $AdS_7\times S^4$ and $\text{Mink}_7\times CY_2$ solutions.

As a result, for the inner product choice of $\mathbb{R}\text{-skew }\xi\eta$ on $M_{11}$, the relation between the solutions $AdS_7\times S^4$ and $\text{Mink}_7\times CY_2$ and the inner product choices on $M_7$ and $M_4$ can be described as in Table VI.

\begin{table}
\centering
\begin{tabular}{c|c}

$M_{11} : \mathbb{R}\text{-skew }\xi\eta$ & $\text{solutions}$\\ \hline
$M_7 : \mathbb{H}^{-}\text{-sym }\xi$ & $\text{Mink}_7\times CY_2$\\
$M_4 : \mathbb{H}\text{-swap }\xi$ & $$\\ \hline
$M_7 : \mathbb{H}^{-}\text{-sym }\xi$ & $\text{Mink}_7\times CY_2$\\
$M_4 : \mathbb{H}^{-}\text{-sym}\oplus\mathbb{H}^{-}\text{-sym }\xi\eta$ & $AdS_7\times S^4$\\ \hline
$M_7 : \mathbb{H}\,\widehat{\,}\text{-sym }\xi\eta$ & $\text{Mink}_7\times CY_2$\\
$M_4 : \mathbb{H}\text{-swap }\xi$ & $$\\ \hline
$M_7 : \mathbb{H}\,\widehat{\,}\text{-sym }\xi\eta$ & $\text{Mink}_7\times CY_2$\\
$M_4 : \mathbb{H}^{-}\text{-sym}\oplus\mathbb{H}^{-}\text{-sym }\xi\eta$ & $AdS_7\times S^4$\\

\end{tabular}
\caption{The relation between the choice of spinor inner products and $M_7\times M_4$ solutions for $\mathbb{R}$-skew $\xi\eta$ inner product on $M_{11}$.}
\end{table}

Note that in the presence of $AdS$ solutions, supergravity Killing forms decompose into special KY forms. For the choice of spinor inner product $\mathbb{R}$-sym $\xi$ on $M$, the supergravity Killing forms $(\epsilon\overline{\epsilon})_p=\{(\epsilon\overline{\epsilon}^{(7)})_p, (\epsilon\overline{\epsilon}^{(4)})_p\}$ satisfy (70) and non-zero bilinear forms are
\begin{eqnarray}
\epsilon\overline{\epsilon}^{(7)}&=&(\epsilon\overline{\epsilon}^{(7)})_0+(\epsilon\overline{\epsilon}^{(7)})_1+(\epsilon\overline{\epsilon}^{(7)})_4+(\epsilon\overline{\epsilon}^{(7)})_5\nonumber\\
\epsilon\overline{\epsilon}^{(4)}&=&(\epsilon\overline{\epsilon}^{(4)})_0+(\epsilon\overline{\epsilon}^{(4)})_1+(\epsilon\overline{\epsilon}^{(4)})_4\nonumber
\end{eqnarray}
on $M_7$ and $M_4$, respectively. For this inner product choice on $M_{11}$, we have the properties of bilinear forms for the chosen inner products on $M_7$ and $M_4$ as given in Table VII.

\begin{table}
\centering
\begin{tabular}{c c|c c c c}

$$ & $\text{inner product}$ & $0$ & $1$ & $4$ & $5$\\ \hline
$i$ & $M_7 : \mathbb{H}^{-}\text{-sym }\xi$ & $R$ & $R$ & $R$ & $R$\\
$$ & $M_4 : \mathbb{H}\text{-swap }\xi$ & $\checkmark$ & $\checkmark$ & $\checkmark$ & $$\\ \hline
$ii$ & $M_7 : \mathbb{H}^{-}\text{-sym }\xi$ & $R$ & $R$ & $R$ & $R$\\
$$ & $M_4 : \mathbb{H}^{-}\text{-sym}\oplus\mathbb{H}^{-}\text{-sym }\xi\eta$ & $R$ & $V$ & $R$ & $$\\ \hline
$iii$ & $M_7 : \mathbb{H}\,\widehat{\,}\text{-sym }\xi\eta$ & $(+)$ & $(-)$ & $(+)$ & $(-)$\\
$$ & $M_4 : \mathbb{H}\text{-swap }\xi$ & $\checkmark$ & $\checkmark$ & $\checkmark$ & $$\\ \hline
$iv$ & $M_7 : \mathbb{H}\,\widehat{\,}\text{-sym }\xi\eta$ & $(+)$ & $(-)$ & $(+)$ & $(-)$\\
$$ & $M_4 : \mathbb{H}^{-}\text{-sym}\oplus\mathbb{H}^{-}\text{-sym }\xi\eta$ & $R$ & $V$ & $R$ & $$\\

\end{tabular}
\caption{Properties of non-zero bilinears for different spinor inner product choices on Lorentzian $M_7$ and Riemannian $M_4$ for $\mathbb{R}$-sym $\xi$ inner product on $M_{11}$.}
\end{table}

We consider four different choices for the decomposition of supergravity Killing forms.

i) $M_7 : \mathbb{H}^{-}\text{-sym }\xi$ and $M_4 : \mathbb{H}\text{-swap }\xi$;

The bilinear form equations for $(\epsilon\overline{\epsilon})_0$, $(\epsilon\overline{\epsilon})_1$, $(\epsilon\overline{\epsilon})_4$, $(\epsilon\overline{\epsilon})_5$, $(\epsilon\overline{\epsilon})_8$ and $(\epsilon\overline{\epsilon})_9$ on $M_7$ can be found as follows
\begin{eqnarray}
\nabla_{X_a}(\epsilon\overline{\epsilon}^{(7)})_0&=&\frac{\lambda}{144}\phi\underset{4}{\wedge}i_{X_a}(\epsilon\overline{\epsilon}^{(7)})_5\nonumber\\
\nabla_{X_a}(\epsilon\overline{\epsilon}^{(7)})_1&=&\frac{\lambda}{144}(e_a\wedge\phi)\underset{4}{\wedge}(\epsilon\overline{\epsilon}^{(7)})_4+\frac{\lambda}{18}(\epsilon\overline{\epsilon}^{(7)})_4\underset{3}{\wedge}i_{X_a}\phi\nonumber\\
\nabla_{X_a}(\epsilon\overline{\epsilon}^{(7)})_4&=&\frac{\lambda}{6}(e_a\wedge\phi)\underset{1}{\wedge}(\epsilon\overline{\epsilon}^{(7)})_1-\frac{\lambda}{36}(e_a\wedge\phi)\underset{3}{\wedge}(\epsilon\overline{\epsilon}^{(7)})_5\nonumber\\
&&+\frac{\lambda}{3}(\epsilon\overline{\epsilon}^{(7)})_1\wedge i_{X_a}\phi+\frac{\lambda}{6}i_{X_a}\phi\underset{2}{\wedge}(\epsilon\overline{\epsilon}^{(7)})_5\\
\nabla_{X_a}(\epsilon\overline{\epsilon}^{(7)})_5&=&\frac{\lambda}{6}e_a\wedge\phi\wedge(\epsilon\overline{\epsilon}^{(7)})_0-\frac{\lambda}{12}(e_a\wedge\phi)\underset{2}{\wedge}(\epsilon\overline{\epsilon}^{(7)})_4-\frac{\lambda}{3}(\epsilon\overline{\epsilon}^{(7)})_4\underset{1}{\wedge}i_{X_a}\phi\nonumber\\
0&=&\frac{\lambda}{6}(e_a\wedge\phi)\underset{1}{\wedge}(\epsilon\overline{\epsilon}^{(7)})_5-\frac{\lambda}{3}i_{X_a}\phi\wedge(\epsilon\overline{\epsilon}^{(7)})_5\nonumber\\
0&=&\frac{\lambda}{6}e_a\wedge\phi\wedge(\epsilon\overline{\epsilon}^{(7)})_4\nonumber
\end{eqnarray}
and on $M_4$, we have
\begin{eqnarray}
\nabla_{X_{\alpha}}(\epsilon\overline{\epsilon}^{(4)})_0&=&0\\
\nabla_{X_{\alpha}}(\epsilon\overline{\epsilon}^{(4)})_1&=&\frac{\mu}{18}(\epsilon\overline{\epsilon}^{(4)})_4\underset{3}{\wedge}i_{X_{\alpha}}z_4\\
\nabla_{X_{\alpha}}(\epsilon\overline{\epsilon}^{(4)})_4&=&\frac{\mu}{3}(\epsilon\overline{\epsilon}^{(4)})_1\wedge i_{X_{\alpha}}z_4.
\end{eqnarray}
The last two equalities of (113) gives $\lambda=0$. From (114), $(\epsilon\overline{\epsilon}^{(4)})_0$ is a constant and from (115) and (116), one finds
\begin{eqnarray}
d(\epsilon\overline{\epsilon}^{(4)})_1&=&0\\
\delta(\epsilon\overline{\epsilon}^{(4)})_1&=&\frac{\mu}{18}(\epsilon\overline{\epsilon}^{(4)})_4\underset{4}{\wedge}z_4
\end{eqnarray}
and
\begin{eqnarray}
d(\epsilon\overline{\epsilon}^{(4)})_4&=&0\\
\delta(\epsilon\overline{\epsilon}^{(4)})_4&=&-\frac{\mu}{3}(\epsilon\overline{\epsilon}^{(4)})_1\underset{1}{\wedge}z_4.
\end{eqnarray}
So, by using the equality (49), one can see that $(\epsilon\overline{\epsilon}^{(4)})_1$ and $(\epsilon\overline{\epsilon}^{(4)})_4$ correspond to CCKY forms
\begin{eqnarray}
\nabla_{X_{\alpha}}(\epsilon\overline{\epsilon}^{(4)})_1&=&-\frac{1}{4}e_{\alpha}\wedge\delta(\epsilon\overline{\epsilon}^{(4)})_1\\
\nabla_{X_{\alpha}}(\epsilon\overline{\epsilon}^{(4)})_4&=&-e_{\alpha}\wedge\delta(\epsilon\overline{\epsilon}^{(4)})_4.
\end{eqnarray}
Moreover, they correspond to special CCKY forms;
\begin{eqnarray}
\nabla_{X_{\alpha}}\delta(\epsilon\overline{\epsilon}^{(4)})_1&=&-\frac{4\mu^2}{9}i_{X_{\alpha}}(\epsilon\overline{\epsilon}^{(4)})_1\\
\nabla_{X_{\alpha}}\delta(\epsilon\overline{\epsilon}^{(4)})_4&=&-\frac{\mu^2}{9}i_{X_{\alpha}}(\epsilon\overline{\epsilon}^{(4)})_4.
\end{eqnarray}
Hence, this inner product choice coresponds to $AdS_7\times S^4$ solution and the geometric Killing spinor $\epsilon_7$ generates the flux component $\phi$ which is a special CCKY form and the geometric Killing spinor $\epsilon_4$ generates the bilinear forms $(\epsilon\overline{\epsilon}^{(4)})_0$, $(\epsilon\overline{\epsilon}^{(4)})_1$ and $(\epsilon\overline{\epsilon}^{(4)})_4$ which are special CCKY forms. If we choose $\mu=0$, then the solution reduces to $\text{Mink}_7\times CY_2$ case and the bilinear forms correspond to parallel forms.

ii) $M_7 : \mathbb{H}^{-}\text{-sym }\xi$ and $M_4 : \mathbb{H}^{-}\text{-sym}\oplus\mathbb{H}^{-}\text{-sym }\xi\eta$;

For that choice of inner products, the bilinear form equations on $M_7$ are the same as in (113) and we have $\lambda=0$. Since, we have the same nonzero bilinears as in case (i), the bilinear form equations on $M_4$ are also the same as in (114)-(116). However, for this choice the bilinear forms $(\epsilon\overline{\epsilon}^{(4)})_0$ and $(\epsilon\overline{\epsilon}^{(4)})_4$ are real quantities while $(\epsilon\overline{\epsilon}^{(4)})_1$ is a vector quaternion. So, the equalities (115) and (116) imply that $\mu$ must be a vector quaterninon. Thus, if we choose the flux 4-form $F$ as a real quantity, then the consistency of (115) and (116) can only be achieved by taking $\mu=0$. Hence, the only solution corresponding to this choice is $\text{Mink}_7\times CY_2$.

iii) $M_7 : \mathbb{H}\,\widehat{\,}\text{-sym }\xi\eta$ and $M_4 : \mathbb{H}\text{-swap }\xi$;

Since all the bilinear forms on $M_7$ are also nonzero in this inner product choice, the equations satisfied by the bilinear forms are the same as in (113). Similarly, the equations on $M_4$ are the same with case (i) and we have $\lambda=0$ and $\mu\neq 0$ with special CCKY forms $(\epsilon\overline{\epsilon}^{(4)})_0$, $(\epsilon\overline{\epsilon}^{(4)})_1$ and $(\epsilon\overline{\epsilon}^{(4)})_4$ on $M_4$ in that case. So, this choice corresponds to $AdS_7\times S^4$ solution and for the special case of $\mu=0$, we have $\text{Mink}_7\times CY_2$ solution.

iv) $M_7 : \mathbb{H}\,\widehat{\,}\text{-sym }\xi\eta$ and $M_4 : \mathbb{H}^{-}\text{-sym}\oplus\mathbb{H}^{-}\text{-sym }\xi\eta$;

The bilinear form equations on $M_7$ corresponds to (113) and the same situation as in case (ii) appears on $M_4$ in this choice and we have $\lambda=\mu=0$. So, the parallel spinors $\epsilon_7$ ad $\epsilon_4$ generate parallel forms and we have $\text{Mink}_7\times CY_2$ solution.

As a result, for the inner product choice of $\mathbb{R}\text{-sym }\xi$ on $M_{11}$, the relation between the solutions $AdS_7\times S^4$ and $\text{Mink}_7\times CY_2$ and the inner product choices on $M_7$ and $M_4$ can be described as in Table VIII.

\begin{table}
\centering
\begin{tabular}{c|c}

$M_{11} : \mathbb{R}\text{-sym }\xi$ & $\text{solutions}$\\ \hline
$M_7 : \mathbb{H}^{-}\text{-sym }\xi$ & $\text{Mink}_7\times CY_2$\\
$M_4 : \mathbb{H}\text{-swap }\xi$ & $AdS_7\times S^4$\\ \hline
$M_7 : \mathbb{H}^{-}\text{-sym }\xi$ & $\text{Mink}_7\times CY_2$\\
$M_4 : \mathbb{H}^{-}\text{-sym}\oplus\mathbb{H}^{-}\text{-sym }\xi\eta$ & $$\\ \hline
$M_7 : \mathbb{H}\,\widehat{\,}\text{-sym }\xi\eta$ & $\text{Mink}_7\times CY_2$\\
$M_4 : \mathbb{H}\text{-swap }\xi$ & $AdS_7\times S^4$\\ \hline
$M_7 : \mathbb{H}\,\widehat{\,}\text{-sym }\xi\eta$ & $\text{Mink}_7\times CY_2$\\
$M_4 : \mathbb{H}^{-}\text{-sym}\oplus\mathbb{H}^{-}\text{-sym }\xi\eta$ & $$\\

\end{tabular}
\caption{The relation between the choice of spinor inner products and $M_7\times M_4$ solutions for $\mathbb{R}$-sym $\xi$ inner product on $M_{11}$.}
\end{table}
 
Note that in the presence of $AdS$ solutions, supergravity Killing forms decompose into special CCKY forms.

\section{$M_5\times M_6$, $M_6\times M_5$ and $M_3\times M_8$ Type Backgrounds}

We can also consider different types of decompositions into product manifolds for $M_{11}$ other than $M_4\times M_7$ and $M_7\times M_4$ decompositions. For example, we can investigate $M_5\times M_6$, $M_6\times M_5$ and $M_3\times M_8$ type backgrounds. Remember that we only consider the unwarped product manifolds and in that case these types of backgrounds will not give interesting examples for the reduction of supergravity Killing form bilinears into KY and CCKY forms by choosing different spinor inner products. The reason for that is the fact that the $AdS$ solutions can only appear for these types of backgrounds in the presence of a warp factor and in the unwarped case we do not have $AdS$ solutions.

In $M_5\times M_6$ case, we have the following decompositions

\begin{eqnarray}
e^A&=&\{e^a\otimes iz_6, 1_5\otimes e^{\alpha}\}\nonumber\\
F&=&\{\lambda\psi, \mu\phi\}\\
\epsilon&=&\epsilon_5\otimes\epsilon_6\nonumber
\end{eqnarray}
where $\psi$ is a 4-form on $M_5$, $\phi$ is a 4-form on $M_6$ and $\lambda$ and $\mu$ are constants. However, for these choices, the consistent decompositions of Maxwell-like, Einstein and supergravity Killing spinor equations can only be possible for the fluxless case $\lambda=0=\mu$. Hence, in that case the solution for all types of spinor inner products is $\text{Mink}_5\times CY_3$.

For $M_6\times M_5$ case, the situation is similar and the only consistent decomposition corresponds to the fluxless case. However, in that case we do not have any solution since there are no five-dimensional compact Riemannian manifolds admitting parallel spinors.

In $M_3\times M_8$, we have the following decompositions
\begin{eqnarray}
e^A&=&\{e^a\otimes z_8, 1_3\otimes e^{\alpha}\}\nonumber\\
F&=&\{0, \mu\phi\}\\
\epsilon&=&\epsilon_3\otimes\epsilon_8\nonumber
\end{eqnarray} 
where $\phi$ is a 4-form on $M_8$. Similarly, the only consistent decomposition is in the fluxless case $\mu=0$ and for all types of spinor inner products the only solution is $\text{Mink}_3\times Spin(7)$. However, in the presence of the warp factor in the decomposition on product manifolds, the field equations will be different and one can find different solutions for the $M_3\times M_8$ decomposition as in \cite{Babalic Lazaroiu}.

\section{Reduction and Lift of KY and CCKY Forms}

The existence of $AdS$ solutions for the unwarped $M_4\times M_7$ and $M_7\times M_4$ type backgrounds is highly dependent on the choice of spinor inner products on product manifolds. As we have have seen in sections III and IV, only some special choices of spinor inner products allow the $AdS$ solutions. Moreoever, we have shown that, for the $AdS$ solutions, there is a relation between supergravity Killing forms on $M_{11}$ and the hidden symmetries on product manifolds. The type of hidden symmetries on product manifolds is also dependent on the choice of the spinor inner product on $M_{11}$. For the choice of $\mathbb{R}$-skew $\xi\eta$ inner product on $M_{11}$, supergravity Killing forms reduce onto special KY 1- and 2-forms on $M_4$. If one chooses $\mathbb{R}$-sym $\xi$ inner product on $M_{11}$, then the supergravity Killing forms reduce onto special CCKY 1- and 4-forms on $M_4$. These are correct for both $M_4\times M_7$ and $M_7\times M_4$ type backgrounds. The situation can be summarized as in Table IX.

\begin{table}
\centering
\begin{tabular}{c | c | c }

inner product on $M_{11}$ & background & reduction of bilinears on $M_4$ \\ \hline
$\mathbb{R}$-skew $\xi\eta$ & $AdS_4\times M_7$ & special KY 1- and 2-forms \\
$\mathbb{R}$-sym $\xi$ & $AdS_4\times M_7$ & special CCKY 1- and 4-forms \\
$\mathbb{R}$-skew $\xi\eta$ & $AdS_7\times S^4$ & special KY 1- and 2-forms  \\
$\mathbb{R}$-sym $\xi$ & $AdS_7\times S^4$ & special CCKY 1- and 4-forms \\

\end{tabular}
\caption{Reduction of bilinears into KY and CCKY forms for $AdS$ solutions depending on the spinor inner product choice on $M_{11}$.}
\end{table}

KY and CCKY forms on product manifolds which are reduced from the supergravity Killing forms on $M_{11}$ can also be lifted to hidden symmetries on $M_{11}$. For any manifold $M$ with a product structure $M=\widetilde{M}_m\times\overline{M}_n$ and metric
\[
g_{AB}=\{\widetilde{g}_{ab}, \overline{g}_{\alpha\beta}\},
\]
one can construct KY and CCKY forms on $M$ by using the KY and CCKY forms on $\overline{M}$. For a KY $p$-form $\overline{\omega}$ on $\overline{M}$ and a CCKY $q$-form $\overline{\nu}$ on $\overline{M}$, the following forms
\begin{eqnarray}
\omega&=&\overline{\omega}\\
\nu&=&z_{\widetilde{M}}\wedge\overline{\nu}
\end{eqnarray}
are KY $p$-forms and CCKY $(m+q)$-forms on $M$, respectively \cite{Krtous Kubiznak Kolar}. Here $z_{\widetilde{M}}$ is the volume form on $\widetilde{M}$. So, for the solutions $AdS_4\times S^7$ and $AdS_4\times\text{weak }G_2$, the internal component of the flux which is the 4-form $\phi$ is a CCKY 4-form and the following form
\begin{equation}
\nu=z_4\wedge\phi
\end{equation}
is a CCKY 8-form on $M_{11}$. For the spinor inner product $\mathbb{R}$-skew $\xi\eta$ on $M_{11}$ and the solution $AdS_7\times S^4$, we have special KY forms $(\epsilon\overline{\epsilon}^{(4)})_1$ and $(\epsilon\overline{\epsilon}^{(4)})_2$ on $S^4$. So, we have the following KY 1- and 2-forms on $M_{11}$
\begin{eqnarray}
\omega_1&=&(\epsilon\overline{\epsilon}^{(4)})_1\nonumber\\
\omega_2&=&(\epsilon\overline{\epsilon}^{(4)})_2.
\end{eqnarray}
However, these do not need to be special KY forms. For the spinor inner product $\mathbb{R}$-sym $\xi$ on $M_{11}$ and the solution $AdS_7\times S^4$, we have the special CCKY forms $(\epsilon\overline{\epsilon}^{(4)})_1$ and $(\epsilon\overline{\epsilon}^{(4)})_4$ on $S^4$. So, we have the following CCKY 8- and 11-forms on $M_{11}$
\begin{eqnarray}
\nu_1&=&z_7\wedge(\epsilon\overline{\epsilon}^{(4)})_1\nonumber\\
\nu_2&=&z_7\wedge(\epsilon\overline{\epsilon}^{(4)})_4.
\end{eqnarray}
Again, these do not need to be special CCKY forms. As a result, supergravity Killing forms constructed out of supergravity Killing spinors induce KY and CCKY forms on $AdS$ backgrounds of eleven-dimensional supergravity.

\section{Conclusion}

We show that the choices of spinor inner products play a central role for the M-theory backgrounds corresponding to unwarped compactifications. Especially, in the case for which the spinor factorizes, the existence of $AdS$ solutions depends on the choice of some special types of spinor inner products on product manifolds. It remains to be determined how these results change when the assumption of spinor factorization is not made. For $AdS$ solutions, supergravity Killing forms which are bilinear forms of supergravity Killing spinors reduce onto the hidden symmetries on product manifolds. These hidden symmetries correspond to special KY and special CCKY forms. Moreover, this reduction gives rise to the lift of hidden symmetries onto eleven-dimensional backgrounds and we find KY and CCKY forms on M-theory backgrounds. The methods leading to the relations between $AdS$ solutions, choices of spinor inner products and reduction to hidden symmetries can be seen as a first step of a classification procedure for general string and M-theory backgrounds in terms of spinor inner products.

One can also investigate the situation for warped product compactifications of M-theory backgrounds. Obviously, the field and bilinear form equations will be different from the unwarped case since they will include the warp factor in that case. On the other hand, these investigations can  also be extended into ten-dimensional string backgrounds and their dependence on the choices of spinor inner products can be determined. So, by finding the relations between solutions, spinor inner products and reduction of bilinear forms, possible classification schemes can be obtained in that way. The conformal field theory equivalent of the choice of spinor inner products can also be investigated in the framework of $AdS/CFT$ correspondence. These may be considered as motivations for future investigations about the topic of the paper.

\begin{acknowledgments}
This work is supported by The Scientific and Technological Research council of Turkey (T\"{U}B\.{I}TAK) Research Project No. 118F086.
\end{acknowledgments}

\begin{appendix}

\section{Inner Product Classes of Spinor Spaces}

\begin{table}
\centering
\begin{tabular}{c|c}

$p-q(\text{mod }8)\quad$ & $Cl_{p,q}$\\ \hline
$0 , 2\quad$ &  $\mathbb{R}(2^{n/2})$\\
$3 , 7\quad$ & $\mathbb{C}(2^{(n-1)/2})$ \\
$4 , 6\quad$ &  $\mathbb{H}(2^{(n-2)/2})$ \\
$1\quad$ &  $\mathbb{R}(2^{(n-1)/2})\oplus\mathbb{R}(2^{(n-1)/2})$ \\
$5\quad$ &  $\mathbb{H}(2^{(n-3)/2})\oplus\mathbb{H}(2^{(n-3)/2})$ \\

\end{tabular}
\caption{Clifford algebras corresponding to $p$ positive and $q$ negative generators}
\end{table}

In this appendix, we will give the possible inner product choices for spinor spaces in different dimensions and signatures. Let us consider the real Clifford algebra $Cl_{p,q}$ with $p$ positive and $q$ negative generators in $n=p+q$ dimensions. It is isomorphic to real, complex or quaternionic matrices as given in Table X. In the table, $\mathbb{D}(k)$ denotes the $k\times k$ matrices with $\mathbb{D}=\mathbb{R},\mathbb{C}$ or $\mathbb{H}$. The even subalgebra $Cl^0_{p,q}$ of a Clifford algebra is isomorphic to a Clifford algebra in one lower dimension as follows
\begin{equation}
Cl^0_{p,q}\cong Cl_{q,p-1}.
\end{equation}
So, we can write the even subalgebras in different dimensions as in Table XI. If we define the spinor spaces as the representation spaces of even subalgebras, then we obtain the classes of spinors in different dimensions as in Table XII. We only consider real Clifford algebras in the paper, but one can also consider complex Clifford algebras, their classification and complex spinors as representations of complex Clifford modules for different investigations.

\begin{table}
\centering
\begin{tabular}{c|c}

$p-q(\text{mod }8)\quad$ & $Cl^0_{p,q}$\\ \hline
$0\quad$ &  $\mathbb{R}(2^{(n-2)/2})\oplus\mathbb{R}(2^{(n-2)/2})$\\
$1 , 7\quad$ & $\mathbb{R}(2^{(n-1)/2})$ \\
$2 , 6\quad$ &  $\mathbb{C}(2^{(n-2)/2})$ \\
$3 , 5\quad$ &  $\mathbb{H}(2^{(n-3)/2})$ \\
$4\quad$ &  $\mathbb{H}(2^{(n-4)/2})\oplus\mathbb{H}(2^{(n-4)/2})$ \\

\end{tabular}
\caption{Even subalgebras of $Cl_{p,q}$.}
\end{table}

\begin{table}
\centering
\begin{tabular}{c | c | c | c}

$p-q(\text{mod }8)$ & $S$ & type of spinors \\ \hline
$0$ & $\mathbb{R}^{2^{(n-2)/2}}\oplus\mathbb{R}^{2^{(n-2)/2}}$ & Majorana-Weyl \\
$1 , 7$ & $\mathbb{R}^{2^{(n-1)/2}}$ & Majorana \\
$2 , 6$ & $\mathbb{C}^{2^{(n-2)/2}}\oplus\mathbb{C}^{2^{(n-2)/2}}$ & Dirac-Weyl  \\
$3 , 5$ & $\mathbb{H}^{2^{(n-3)/2}}$ & Symplectic Majorana \\
$4$ & $\mathbb{H}^{2^{(n-4)/2}}\oplus\mathbb{H}^{2^{(n-4)/2}}$ & Symplectic Majorana-Weyl \\

\end{tabular}
\caption{Spinor spaces $S$ and the classes of spinors for different $p$ and $q$ values.}
\end{table}

One can define different types of inner products $(\,,\,)$ on representation spaces of Clifford algebras. If $\psi$ and $\phi$ are elements of representation spaces of Clifford algebras, then we have
\begin{equation}
(\psi, \phi)=\pm(\phi, \psi)^j
\end{equation}
 which are called $\mathbb{D}^j$-symmetric or $\mathbb{D}^j$-skew inner products respectively where $j$ denotes the identity for $\mathbb{D}=\mathbb{R}$, identity or complex conjugation ($^*$) for $\mathbb{D}=\mathbb{C}$, quaternionic conjugation $(\,^{\overline{\,}}\,)$ or quaterninonic reversion $(\,^{\widehat{\,}}\,)$ for $\mathbb{D}=\mathbb{H}$. Moreover, for any Clifford form $\omega$, we have the following property
\begin{equation}
(\psi, \omega.\phi)=(\omega^{\mathcal{J}}.\psi, \phi)
\end{equation}
where $\mathcal{J}$ corresponds to $\xi$ or $\xi\eta$ involutions on the Clifford algebra and $.$ denotes the Clifford product which is defined as in (B1) and (B2). Here $\xi$ denotes the anti-involution acting on any $p$-form $\omega$ as $\omega^{\xi}=(-1)^{\lfloor p/2\rfloor}\omega$ and $\eta$ is the inner automorphism acting as $\omega^{\eta}=(-1)^p\omega$. $\lfloor\,\rfloor$ denotes the floor function which takes the integer part of the argument. So, we have three choices for an inner product; symmetry or anti-symmetry, the involution $\mathcal{J}$ and the induced involution $j$. From the detailed analysis of Clifford algebras, one can see that there are ten different types of inner products on real Clifford algebras as in Table XIII \cite{Benn Tucker}. In the table, swap means that when the arguments in the inner product are reversed, their semi-spinor space is changed.The inner products induced on Clifford algebra representations in different dimensions can be listed as in Table XIV \cite{Benn Tucker}. In the table, for each dimension, the first row corresponds to the inner product with $\xi$ involution and the second row corresponds to the inner product with $\xi\eta$ involution and the numbers in the table corresponds to the inner product classes in the table XIII. The inner product classes $k\oplus k$ denotes $k$th inner product class on each semi-spinor space. The table repeats itself after dimension 7 with respect to mod 8. As the representation spaces of even subalgebras, the inner products on spinor spaces can also be obtained from Table XIV via the isomorphism $Cl^0_{p,q}\cong Cl_{q,p-1}$ as in Table XV.

\begin{table}
\centering
\begin{tabular}{c | c c c | c}

$1$ & $\mathbb{R}$-sym & \quad\quad & $6$ & $\mathbb{H}^-$-sym \\
$2$ & $\mathbb{R}$-skew & \quad\quad & $7$ & $\mathbb{H}^{\widehat{\,\,}}$-sym \\
$3$ & $\mathbb{C}$-sym & \quad\quad & $8$ & $\mathbb{R}$-swap \\
$4$ & $\mathbb{C}$-skew & $\quad\quad$ & $9$ & $\mathbb{H}$-swap \\
$5$ & $\mathbb{C}^*$-sym & \quad\quad & $10$ & $\mathbb{C}$-swap \\

\end{tabular}
\caption{Types of inner products for real Clifford algebras.}
\end{table}

\begin{table}
\centering
\begin{tabular}{c | c c c c c c c c | c}

$Cl_{p,q}$ & $0$ & $1$ & $2$ & $3$ & $4$ & $5$ & $6$ & $7$ & $\mathcal{J}$ \\ \hline
$\,$ & $\,$ & $\,$ & $\,$ & $\,$ & $\,$ & $\,$ & $\,$ & $\,$ & $\,$ \\
$0$ & $1$ & $3$ & $7$ & $9$ & $6$ & $4$ & $2$ & $8$ & $\xi$ \\
$\,$ & $1$ & $5$ & $6$ & $6\oplus 6$ & $6$ & $5$ & $1$ & $1\oplus 1$ & $\xi\eta$ \\
$\,$ & $\,$ & $\,$ & $\,$ & $\,$ & $\,$ & $\,$ & $\,$ & $\,$ & $\,$ \\
$1$ & $1\oplus 1$ & $1$ & $5$ & $6$ & $6\oplus 6$ & $6$ & $5$ & $1$ & $\xi$ \\
$\,$ & $8$ & $2$ & $4$ & $6$ & $9$ & $7$ & $3$ & $1$ & $\xi\eta$ \\
$\,$ & $\,$ & $\,$ & $\,$ & $\,$ & $\,$ & $\,$ & $\,$ & $\,$ & $\,$ \\
$2$ & $1$ & $8$ & $2$ & $4$ & $6$ & $9$ & $7$ & $3$ & $\xi$ \\
$\,$ & $2$ & $2\oplus 2$ & $2$ & $5$ & $7$ & $7\oplus 7$ & $7$ & $5$ & $\xi\eta$ \\
$\,$ & $\,$ & $\,$ & $\,$ & $\,$ & $\,$ & $\,$ & $\,$ & $\,$ & $\,$ \\
$3$ & $5$ & $2$ & $2\oplus 2$ & $2$ & $5$ & $7$ & $7\oplus 7$ & $7$ & $\xi$ \\
$\,$ & $4$ & $2$ & $8$ & $1$ & $3$ & $7$ & $9$ & $6$ & $\xi\eta$ \\
$\,$ & $\,$ & $\,$ & $\,$ & $\,$ & $\,$ & $\,$ & $\,$ & $\,$ & $\,$ \\
$4$ & $6$ & $4$ & $2$ & $8$ & $1$ & $3$ & $7$ & $9$ & $\xi$ \\
$\,$ & $6$ & $5$ & $1$ & $1\oplus 1$ & $1$ & $5$ & $6$ & $6\oplus 6$ & $\xi\eta$ \\
$\,$ & $\,$ & $\,$ & $\,$ & $\,$ & $\,$ & $\,$ & $\,$ & $\,$ & $\,$ \\
$5$ & $6\oplus 6$ & $6$ & $5$ & $1$ & $1\oplus 1$ & $1$ & $5$ & $6$ & $\xi$ \\
$\,$ & $9$ & $7$ & $3$ & $1$ & $8$ & $2$ & $4$ & $6$ & $\xi\eta$ \\
$\,$ & $\,$ & $\,$ & $\,$ & $\,$ & $\,$ & $\,$ & $\,$ & $\,$ & $\,$ \\
$6$ & $6$ & $9$ & $7$ & $3$ & $1$ & $8$ & $2$ & $4$ & $\xi$ \\
$\,$ & $7$ & $7\oplus 7$ & $7$ & $5$ & $2$ & $2\oplus 2$ & $2$ & $5$ & $\xi\eta$ \\
$\,$ & $\,$ & $\,$ & $\,$ & $\,$ & $\,$ & $\,$ & $\,$ & $\,$ & $\,$ \\
$7$ & $5$ & $7$ & $7\oplus 7$ & $7$ & $2$ & $2$ & $2\oplus 2$ & $2$ & $\xi$ \\
$\,$ & $3$ & $7$ & $9$ & $6$ & $4$ & $2$ & $8$ & $1$ & $\xi\eta$ \\

\end{tabular}
\caption{The inner products induced on Clifford algebra representations for different dimensions where the rows denote the positive generators $p$ and the columns denote the negative generators $q$.}
\end{table}

\begin{table}
\centering
\begin{tabular}{c | c c c c c c c c | c}

$Cl^0_{p,q}$ & $0$ & $1$ & $2$ & $3$ & $4$ & $5$ & $6$ & $7$ & $\mathcal{J}$ \\ \hline
$\,$ & $\,$ & $\,$ & $\,$ & $\,$ & $\,$ & $\,$ & $\,$ & $\,$ & $\,$ \\
$0$ & $8$ & $1$ & $3$ & $7$ & $9$ & $6$ & $4$ & $2$ & $\xi$ \\
$\,$ & $1\oplus 1$ & $1$ & $5$ & $6$ & $6\oplus 6$ & $6$ & $5$ & $1$ & $\xi\eta$ \\
$\,$ & $\,$ & $\,$ & $\,$ & $\,$ & $\,$ & $\,$ & $\,$ & $\,$ & $\,$ \\
$1$ & $1$ & $1\oplus 1$ & $1$ & $5$ & $6$ & $6\oplus 6$ & $6$ & $5$ & $\xi$ \\
$\,$ & $1$ & $8$ & $2$ & $4$ & $6$ & $9$ & $7$ & $3$ & $\xi\eta$ \\
$\,$ & $\,$ & $\,$ & $\,$ & $\,$ & $\,$ & $\,$ & $\,$ & $\,$ & $\,$ \\
$2$ & $3$ & $1$ & $8$ & $2$ & $4$ & $6$ & $9$ & $7$ & $\xi$ \\
$\,$ & $5$ & $2$ & $2\oplus 2$ & $2$ & $5$ & $7$ & $7\oplus 7$ & $7$ & $\xi\eta$ \\
$\,$ & $\,$ & $\,$ & $\,$ & $\,$ & $\,$ & $\,$ & $\,$ & $\,$ & $\,$ \\
$3$ & $7$ & $5$ & $2$ & $2\oplus 2$ & $2$ & $5$ & $7$ & $7\oplus 7$ & $\xi$ \\
$\,$ & $6$ & $4$ & $2$ & $8$ & $1$ & $3$ & $7$ & $9$ & $\xi\eta$ \\
$\,$ & $\,$ & $\,$ & $\,$ & $\,$ & $\,$ & $\,$ & $\,$ & $\,$ & $\,$ \\
$4$ & $9$ & $6$ & $4$ & $2$ & $8$ & $1$ & $3$ & $7$ & $\xi$ \\
$\,$ & $6\oplus 6$ & $6$ & $5$ & $1$ & $1\oplus 1$ & $1$ & $5$ & $6$ & $\xi\eta$ \\
$\,$ & $\,$ & $\,$ & $\,$ & $\,$ & $\,$ & $\,$ & $\,$ & $\,$ & $\,$ \\
$5$ & $6$ & $6\oplus 6$ & $6$ & $5$ & $1$ & $1\oplus 1$ & $1$ & $5$ & $\xi$ \\
$\,$ & $6$ & $9$ & $7$ & $3$ & $1$ & $8$ & $2$ & $4$ & $\xi\eta$ \\
$\,$ & $\,$ & $\,$ & $\,$ & $\,$ & $\,$ & $\,$ & $\,$ & $\,$ & $\,$ \\
$6$ & $4$ & $6$ & $9$ & $7$ & $3$ & $1$ & $8$ & $2$ & $\xi$ \\
$\,$ & $5$ & $7$ & $7\oplus 7$ & $7$ & $5$ & $2$ & $2\oplus 2$ & $2$ & $\xi\eta$ \\
$\,$ & $\,$ & $\,$ & $\,$ & $\,$ & $\,$ & $\,$ & $\,$ & $\,$ & $\,$ \\
$7$ & $2$ & $5$ & $7$ & $7\oplus 7$ & $7$ & $5$ & $2$ & $2\oplus 2$ & $\xi$ \\
$\,$ & $1$ & $3$ & $7$ & $9$ & $6$ & $4$ & $2$ & $8$ & $\xi\eta$ \\

\end{tabular}
\caption{The inner products induced on even subalgebra representations for different dimensions where the rows denote the positive generators $p$ and the columns denote the negative generators $q$.}
\end{table}

On a spin manifold $M$, the possible inner product choices on the spinor bundle can be determined from the Table XV. So, the manifolds that we consider throughout the text can have the spinor inner products given in Table XVI. These inner products correspond to the admissible inner products analyzed in \cite{Alekseevsky Cortes, Alekseevsky Cortes Devchand, Alekseevsky Cortes Devchand Proeyen}. The conditions for a spinor inner product to be admissible are as follows. For a 1-form $v$ and spinors $\psi$ and $\phi$, the inner product $(\,,\,)$ satisfies
\[
(v.\psi,\phi)=\tau(\psi,v.\phi)
\]
where $\tau$ corresponds to the type of the inner product and equals to $+1$ or $-1$. The inner product is symmetric or skew-symmetric
\[
(\psi,\phi)=\sigma(\phi,\psi)
\]
where $\sigma$ corresponds to the symmetry of the inner product and equals to $+1$ or $-1$, respectively. If the spinor space is reducible to semi-spinor spaces, then the semi-spinor spaces are mutually orthogonal or isotropic which is determined by the isotropy $\iota$ of the inner product and equals to $+1$ or $-1$, respectively. The classification of admissible inner products in all dimensions and signatures given in \cite{Alekseevsky Cortes, Alekseevsky Cortes Devchand Proeyen} are equivalent to the classification given in \cite{Benn Tucker}. The type $\tau=\pm1$ of the inner product corresponds to the choice of involution ${\mathcal{J}}=\xi$ or $\xi\eta$. The symmetry $\sigma=\pm1$ corresponds to the symmetric or skew inner products and isotropy $\iota=\pm1$ corresponds to the sym$\oplus$sym or swap inner products on semi-spinor spaces. Since the classification of spinor inner products given in \cite{Benn Tucker} is more compact for all dimensions and signatures, we prefer to use it in the paper.

\begin{table}
\centering
\begin{tabular}{c | c c c}

$\text{11-d Lorentzian}$ & $\mathbb{R}\text{-sym }\xi$ & $\,,$ & $\mathbb{R}\text{-skew }\xi\eta$ \\
$\text{3-d Lorentzian}$ & $\mathbb{R}\text{-sym }\xi$ & $\,,$ & $\mathbb{R}\text{-skew }\xi\eta$ \\
$\text{4-d Lorentzian}$ & $\mathbb{C}^*\text{-sym }\xi$ & $\,,$ & $\mathbb{C}\text{-skew }\xi\eta$ \\
$\text{5-d Lorentzian}$ & $\mathbb{H}^{-}\text{-sym }\xi$ & $\,,$ & $\mathbb{H}^{-}\text{-sym }\xi\eta$ \\
$\text{6-d Lorentzian}$ & $\mathbb{H}^{-}\text{-sym}\oplus\mathbb{H}^{-}\text{-sym }\xi$ & $\,,$ & $\mathbb{H}\text{-swap }\xi\eta$ \\
$\text{7-d Lorentzian}$ & $\mathbb{H}^{-}\text{-sym }\xi$ & $\,,$ & $\mathbb{H}\,\widehat{\,}\text{-sym }\xi\eta$ \\
$\text{8-d Riemannian}$ & $\mathbb{R}\text{-swap }\xi$ & $\,,$ & $\mathbb{R}\text{-sym}\oplus\mathbb{R}\text{-sym }\xi\eta$ \\
$\text{7-d Riemannian}$ & $\mathbb{R}\text{-skew }\xi$ & $\,,$ & $\mathbb{R}\text{-sym }\xi\eta$ \\
$\text{6-d Riemannian}$ & $\mathbb{C}\text{-skew }\xi$ & $\,,$ & $\mathbb{C}^*\text{-sym }\xi\eta$ \\
$\text{5-d Riemannian}$ & $\mathbb{H}^{-}\text{-sym }\xi$ & $\,,$ & $\mathbb{H}^{-}\text{-sym }\xi\eta$ \\
$\text{4-d Riemannian}$ & $\mathbb{H}\text{-swap }\xi$ & $\,,$ & $\mathbb{H}^{-}\text{-sym}\oplus\mathbb{H}^{-}\text{sym }\xi\eta$ \\

\end{tabular}
\caption{Possible spinor inner products for the manifolds used in the text.}
\end{table}

For any spinor field $\epsilon$, the choice of the inner product determines the properties of the bilinear forms constructed from $\epsilon$. For a $p$-form $\omega$, we have
\begin{equation}
(\epsilon, \omega.\epsilon)=\pm(\epsilon, \omega^{\mathcal{J}}.\epsilon)^j
\end{equation}
and if we take $\omega$ as the $p$-form basis, then symmetry or antisymmetry  of the inner product and the choice of involution $\mathcal{J}$ determine the properties of the bilinear $p$-form. For example, on an eleven-dimensional Lorentzian manifold $M_{11}$ with the spinor inner product $\mathbb{R}$-skew $\xi\eta$, the 3-form bilinear corresponds to
\begin{eqnarray}
(\epsilon, (e^c\wedge e^b\wedge e^a).\epsilon)&=&-((e^c\wedge e^b\wedge e^a).\epsilon, \epsilon)\nonumber\\
&=&-(\epsilon, (e^c\wedge e^b\wedge e^a)^{\xi\eta}.\epsilon)\nonumber\\
&=&-(\epsilon, (e^c\wedge e^b\wedge e^a).\epsilon)\nonumber
\end{eqnarray}
which means that it vanishes automatically. However, for a 2-form bilinear, we have
\begin{eqnarray}
(\epsilon, (e^b\wedge e^a).\epsilon)&=&-((e^b\wedge e^a).\epsilon, \epsilon)\nonumber\\
&=&-(\epsilon, (e^b\wedge e^a)^{\xi\eta}.\epsilon)\nonumber\\
&=&(\epsilon, (e^b\wedge e^a).\epsilon)\nonumber
\end{eqnarray}
and hence it is nonzero. For the inner product choices of the manifolds that are considered in the text, the properties of bilinear $p$-forms for different form degrees can be summarized as in Table XVII. When the induced involution $j$ is the identity, some of the bilinear forms vanish and these are denoted by $\times$ in the table while the non-vanishing ones are denoted by $\checkmark$. When $j$ is the complex conjugation, the bilinear forms are real or pure imaginary and these are denoted in the table as $R$ and $I$, respectively. When $j$ is the quaternionic conjugation, the bilinear forms are real or vector quaternions and these are denoted in the table as $R$ and $V$, respectively. For the case of $j$ corresponding to the quaternionic reversion, the bilinear forms are symmetric or antisymmetric under reversion operation and these are denoted in the table as $(+)$ and $(-)$, respectively. The properties of the bilinear forms resulting from the decompositions of eleven-dimensional supergravity backgrounds in the main text can be deduced from this table.

\begin{table}
\centering
\begin{tabular}{c | c | c c c c c c c c c c c c}

$\text{manifold}$ & $\text{inner product}$ & $0$ & $1$ & $2$ & $3$ & $4$ & $5$ & $6$ & $7$ & $8$ & $9$ & $10$ & $11$ \\ \hline
$\text{Lorentzian }M_{11}$ & $\mathbb{R}\text{-skew }\xi\eta$ & $\times$ & $\checkmark$ & $\checkmark$ & $\times$ & $\times$ & $\checkmark$ & $\checkmark$ & $\times$ & $\times$ & $\checkmark$ & $\checkmark$ & $\times$ \\
$\text{Lorentzian }M_{11}$ & $\mathbb{R}\text{-sym }\xi$ & $\checkmark$ & $\checkmark$ & $\times$ & $\times$ & $\checkmark$ & $\checkmark$ & $\times$ & $\times$ & $\checkmark$ & $\checkmark$ & $\times$ & $\times$ \\
$\text{Lorentzian }M_{3}$ & $\mathbb{R}\text{-sym }\xi$ & $\checkmark$ & $\checkmark$ & $\times$ & $\times$ & $$ & $$ & $$ & $$ & $$ & $$ & $$ & $$ \\
$\text{Lorentzian }M_{3}$ & $\mathbb{R}\text{-skew }\xi\eta$ & $\times$ & $\checkmark$ & $\checkmark$ & $\times$ & $$ & $$ & $$ & $$ & $$ & $$ & $$ & $$ \\
$\text{Lorentzian }M_{4}$ & $\mathbb{C}^*\text{-sym }\xi$ & $R$ & $R$ & $I$ & $I$ & $R$ & $$ & $$ & $$ & $$ & $$ & $$ & $$ \\
$\text{Lorentzian }M_{4}$ & $\mathbb{C}\text{-skew }\xi\eta$ & $\times$ & $\checkmark$ & $\checkmark$ & $\times$ & $\times$ & $$ & $$ & $$ & $$ & $$ & $$ & $$ \\
$\text{Lorentzian }M_{5}$ & $\mathbb{H}^{-}\text{-sym }\xi$ & $R$ & $R$ & $V$ & $V$ & $R$ & $R$ & $$ & $$ & $$ & $$ & $$ & $$ \\
$\text{Lorentzian }M_{5}$ & $\mathbb{H}^{-}\text{-sym }\xi\eta$ & $R$ & $V$ & $V$ & $R$ & $R$ & $V$ & $$ & $$ & $$ & $$ & $$ & $$ \\
$\text{Lorentzian }M_{6}$ & $\mathbb{H}^{-}\text{-sym}\oplus\mathbb{H}^{-}\text{-sym }\xi$ & $R$ & $R$ & $V$ & $V$ & $R$ & $R$ & $V$ & $$ & $$ & $$ & $$ & $$ \\
$\text{Lorentzian }M_{6}$ & $\mathbb{H}\text{-swap }\xi\eta$ & $\checkmark$ & $\times$ & $\times$ & $\checkmark$ & $\checkmark$ & $\times$ & $\times$ & $$ & $$ & $$ & $$ & $$ \\
$\text{Lorentzian }M_{7}$ & $\mathbb{H}^{-}\text{-sym }\xi$ & $R$ & $R$ & $V$ & $V$ & $R$ & $R$ & $V$ & $V$ & $$ & $$ & $$ & $$ \\
$\text{Lorentzian }M_{7}$ & $\mathbb{H}\,\widehat{\,}\text{-sym }\xi\eta$ & $(+)$ & $(-)$ & $(-)$ & $(+)$ & $(+)$ & $(-)$ & $(-)$ & $(+)$ & $$ & $$ & $$ & $$ \\
$\text{Riemannian }M_{8}$ & $\mathbb{R}\text{-swap }\xi$ & $\checkmark$ & $\checkmark$ & $\times$ & $\times$ & $\checkmark$ & $\checkmark$ & $\times$ & $\times$ & $\checkmark$ & $$ & $$ & $$ \\
$\text{Riemannian }M_{8}$ & $\mathbb{R}\text{-sym}\oplus\mathbb{R}\text{-sym }\xi\eta$ & $\checkmark$ & $\times$ & $\times$ & $\checkmark$ & $\checkmark$ & $\times$ & $\times$ & $\checkmark$ & $\checkmark$ & $$ & $$ & $$ \\
$\text{Riemannian }M_{7}$ & $\mathbb{R}\text{-skew }\xi$ & $\times$ & $\times$ & $\checkmark$ & $\checkmark$ & $\times$ & $\times$ & $\checkmark$ & $\checkmark$ & $$ & $$ & $$ & $$ \\
$\text{Riemannian }M_{7}$ & $\mathbb{R}\text{-sym }\xi\eta$ & $\checkmark$ & $\times$ & $\times$ & $\checkmark$ & $\checkmark$ & $\times$ & $\times$ & $\checkmark$ & $$ & $$ & $$ & $$ \\
$\text{Riemannian }M_{6}$ & $\mathbb{C}\text{-skew }\xi$ & $\times$ & $\times$ & $\checkmark$ & $\checkmark$ & $\times$ & $\times$ & $\checkmark$ & $$ & $$ & $$ & $$ & $$ \\
$\text{Riemannian }M_{6}$ & $\mathbb{C}^*\text{-sym }\xi\eta$ & $R$ & $I$ & $I$ & $R$ & $R$ & $I$ & $I$ & $$ & $$ & $$ & $$ & $$ \\
$\text{Riemannian }M_{5}$ & $\mathbb{H}^{-}\text{-sym }\xi$ & $R$ & $R$ & $V$ & $V$ & $R$ & $R$ & $$ & $$ & $$ & $$ & $$ & $$ \\
$\text{Riemannian }M_{5}$ & $\mathbb{H}^{-}\text{-sym }\xi\eta$ & $R$ & $V$ & $V$ & $R$ & $R$ & $V$ & $$ & $$ & $$ & $$ & $$ & $$ \\
$\text{Riemannian }M_{4}$ & $\mathbb{H}\text{-swap }\xi$ & $\checkmark$ & $\checkmark$ & $\times$ & $\times$ & $\checkmark$ & $$ & $$ & $$ & $$ & $$ & $$ & $$ \\
$\text{Riemannian }M_{4}$ & $\mathbb{H}^{-}\text{-sym}\oplus\mathbb{H}^{-}\text{-sym }\xi\eta$ & $R$ & $V$ & $V$ & $R$ & $R$ & $$ & $$ & $$ & $$ & $$ & $$ & $$ \\

\end{tabular}
\caption{The properties of bilinear $p$-forms for the inner product choices of the manifolds that are considered in the text where the columns correspond to the value of $p$.}
\end{table}

In general, only the $\xi\eta$ involution inner products on spinor spaces in odd dimensions are considered in supergravity literature. Since the spinors are elements of the even subalgebra and the involution $\eta$ acts trivially on even elements ($\omega^{\eta}=\omega$ for a $2p$-form $\omega$), the effects of the involutions $\xi$ and $\xi\eta$ on spinors are equivalent and taking one of them as an involution is enough to define an inner product. This will be relevant if the only variable in the definition of the spinor inner product would be the definition of spinor duality in terms of the involutions; $\overline{\psi}=\psi^{\mathcal{J}}=\psi^{\xi}=\psi^{\xi\eta}$ and this would lead to a unique bilinear construction in odd dimensions. However, in the definition of spinor inner products, the involutions also act on Clifford forms acting on spinors as defined in (A3) and these Clifford forms can be even or odd. Hence, the effect of involutions $\xi$ and $\xi\eta$ are different on different degree Clifford forms and this gives different inner product definitions for different choices of involutions. Namely, we can consider another inner product different from the standard $\xi\eta$ one and check whether it also correspond to physical solutions. This is what we have done in the paper. Moreover, choosing different involutions in the definition of inner products will also effect the properties of bilinears as can be seen from (A4), but those different properties of bilinears will still give the same physical conclusions. For example, on an 11-dimensional Lorentzian manifold $M_{11}$, the standard physical spinor inner product is chosen as $\mathbb{R}$-skew $\xi\eta$ and all bilinears are calculated with respect to this inner product. However, as discussed above, there is one more possibility for the inner product which corresponds to $\mathbb{R}$-sym $\xi$ which gives different types of non-zero bilinears.

For example, in Chapter 3, for $M_4\times M_7$ decomposition we first consider physically relevant (supersymmetric) inner product $\mathbb{R}$-skew $\xi\eta$ and we find, as expected, the physical (Minkowski and $AdS$) solutions when we choose $\xi\eta$ involution inner products in odd dimensions and both involutions in even dimensions (see Table II). Then, we consider the second possibility $\mathbb{R}$-sym $\xi$ for the inner product (corresponding to non-supersymmetric case) and check whether we still can find physical solutions. As can be seen from Table IV, if we take both involutions in odd dimensions and $\xi$ involution in even dimensions, we can obtain physical (both Minkowski and $AdS$) solutions. This shows that without \textit{a priori} choices for the inner product, one can find physical solutions by considering all possible inner products as expected. However, this does not mean that we can choose inner products completely arbitrarily since some choices give only mathematical solutions. In fact, the inner product choices corresponding to $AdS$ solutions that are given in the Tables II, IV, VI and VIII correspond to "physical" inner product choices. Moreover, the reduction of bilinears onto product manifolds also differ for different choices of inner products. For the case of $\mathbb{R}$-skew $\xi\eta$ inner product, 11-dimensional bilinears will reduce to special KY 1- and 2-forms and for $\mathbb{R}$-sym $\xi$ inner product, they will reduce to special CCKY 1- and 4-forms on 4-dimensional product manifolds (see Table IX). In literature, only the Killing vector properties of 1-form bilinears are mentioned and hidden symmetry properties of higher form bilinears are not discussed since only one choice of inner product is considered. So, we extend the relevant inner product classes for physical solutions by considering all mathematically possible inner product choices and give a detailed account about the types of possible inner product choices and the corresponding types of solutions (physical or other) without \textit{a priori} assumptions.

\section{Clifford algebra conventions and Clifford bracket}

On the exterior bundle $\Lambda M$ on a $n$-dimensional manifold $M$, besides the wedge product $\wedge$, one can also define the Clifford product $.$ . This turns $\Lambda M$ into a Clifford bundle $Cl(M)$ on $M$. Sections of $Cl(M)$ are called Clifford forms. On the Clifford bundle, the coframe basis $\{e^a\}$ satisfy the following Clifford algebra identity
\begin{equation}
e^a.e^b+e^b.e^a=2g^{ab}.
\end{equation}
where $g^{ab}$ is the inverse metric. The Clifford product can be written in terms of the wedge product and interior derivative. For any $p$-form $\omega$, we have the following identities
\begin{eqnarray}
e^a.\omega&=&e^a\wedge\omega+i_{X^a}\omega\nonumber\\
\omega.e^a&=&e^a\wedge\eta\omega-i_{X^a}\eta\omega
\end{eqnarray}
and similarly the wedge product and interior derivative can be written in terms of the Clifford product as
\begin{eqnarray}
e^a\wedge\omega&=&\frac{1}{2}(e^a.\omega+\eta\omega.e^a)\nonumber\\
i_{X^a}\omega&=&\frac{1}{2}(e^a.\omega-\eta\omega.e^a).
\end{eqnarray}
From these equalities, one can also deduce that
\begin{equation}
e^a.\omega.e_a=(n-2p)\eta\omega.
\end{equation}
For any two Clifford forms $\alpha$ and $\beta$ which correspond to inhomogeneous differential forms, one can write the Clifford product from (B2) as in the following form
\begin{equation}
\alpha.\beta=\sum_{k=0}^n\frac{(-1)^{\lfloor k/2\rfloor}}{k!}(\eta^ki_{X_{a_1}}i_{X_{a_2}}...i_{X_{a_k}}\alpha)\wedge(i_{X^{a_1}}i_{X^{a_2}}...i_{X^{a^k}}\beta).
\end{equation}
Moreover, we can also define the Clifford commutator $[\,,\,]_{Cl}$ as
\begin{equation}
[\alpha, \beta]_{Cl}=\alpha.\beta-\beta.\alpha
\end{equation}
and from (B5), it can be written as
\begin{eqnarray}
[\alpha, \beta]_{Cl}=\sum_{k=0}^n\frac{(-1)^{\lfloor k/2\rfloor}}{k!}&&\bigg[(\eta^ki_{X_{a_1}}i_{X_{a_2}}...i_{X_{a_k}}\alpha)\wedge(i_{X^{a_1}}i_{X^{a_2}}...i_{X^{a^k}}\beta)\nonumber\\
&&-(\eta^ki_{X_{a_1}}i_{X_{a_2}}...i_{X_{a_k}}\beta)\wedge(i_{X^{a_1}}i_{X^{a_2}}...i_{X^{a_k}}\alpha)\bigg].
\end{eqnarray}
To write it in a more compact form, we define the contracted wedge product
\begin{equation}
\alpha\underset{k}{\wedge}\beta=i_{X_{a_1}}i_{X_{a_2}}...i_{X_{a_k}}\alpha\wedge i_{X^{a_1}}i_{X^{a_2}}...i_{X^{a^k}}\beta
\end{equation}
and (B7) turns into
\begin{equation}
[\alpha, \beta]_{Cl}=\sum_{k=0}^n\frac{(-1)^{\lfloor k/2\rfloor}}{k!}\left[\eta^k\alpha\underset{k}{\wedge}\beta-\eta^k\beta\underset{k}{\wedge}\alpha\right].
\end{equation}
For example, if we consider the special case where $\alpha$ is a 2-form and $\beta$ arbitrary, then the right hand side of the Clifford commutator only has one nonzero term and we have
\begin{equation}
[\alpha, \beta]_{Cl}=-2\alpha\underset{1}{\wedge}\beta.
\end{equation}
If we apply the projection operator $(\,)_p$ to the Clifford commutator, that is $\left([\alpha, \beta]_{Cl}\right)_p$, then it gives only the $p$-form part of the right hand side of (B9). Similarly, we can define the Clifford anticommutator as
\begin{equation}
[\alpha, \beta]_{+Cl}=\alpha.\beta+\beta.\alpha
\end{equation}
and from (B5), it can be written as
\begin{eqnarray}
[\alpha, \beta]_{+Cl}=\sum_{k=0}^n\frac{(-1)^{\lfloor k/2\rfloor}}{k!}\left[\eta^k\alpha\underset{k}{\wedge}\beta+\eta^k\beta\underset{k}{\wedge}\alpha\right].
\end{eqnarray}

\section{Killing-Yano forms}

Killing vector fields correspond to the symmetries of a manifold and the antisymmetric generalizations of them to the higher-degree differential forms are Killing-Yano (KY) forms which are called the hidden symmetries of the manifold. A $p$-form $\omega$ is a KY $p$-form if it satisfies the following equation
\begin{equation}
\nabla_X\omega=\frac{1}{p+1}i_Xd\omega
\end{equation}
for any vector field $X$. Similarly, conformal Killing vector fields can also be generalized to higher-degree differential forms and those are called conformal Killing-Yano (CKY) forms. A $p$-form $\omega$ is a CKY $p$-form on a $n$-dimensional manifold $M$, if it satisfies the following equation
\begin{equation}
\nabla_X\omega=\frac{1}{p+1}i_Xd\omega-\frac{1}{n-p+1}\widetilde{X}\wedge\delta\omega
\end{equation}
for any vector field $X$. So, KY forms correspond to coclosed CKY forms satisfying $\delta\omega=0$. Another subset of CKY forms satisfiying $d\omega=0$ are called closed conformal Killing-Yano (CCKY) forms and hence they are solutions of the following equation
\begin{equation}
\nabla_X\omega=-\frac{1}{n-p+1}\widetilde{X}\wedge\delta\omega.
\end{equation}
We can also define special subsets of the spaces of KY and CCKY forms. A KY $p$-form $\omega$ is called a special KY $p$-form if it satisfies the following condition
\begin{equation}
\nabla_Xd\omega=-c(p+1)\widetilde{X}\wedge\omega
\end{equation}
for a constant $c$ \cite{Semmelmann}. Similarly, a CCKY $p$-form $\omega$ is called a special CCKY $p$-form if it satisfies the following condition
\begin{equation}
\nabla_X\delta\omega=c(n-p+1)i_X\omega.
\end{equation}
The importance of the special KY and CCKY forms is the fact that they have to be generated from geometric Killing spinors as bilinear forms \cite{Acik Ertem1}. Non-special KY and CCKY forms cannot be generated by geometric Killing spinors. 

\end{appendix}


\end{document}